\documentclass[twocolumn,
               secnumarabic,
               amssymb,
               nobibnotes,
               aps,
               prm,
               superscriptaddress
               ]{revtex4-1}
\usepackage{bm}
\usepackage{graphicx}
\graphicspath{{images/}}

\usepackage{amsmath}
\usepackage{mathtools}
\usepackage{textcomp}

\begin{document}

\title{Pinning Dislocations in Colloidal Crystals with Active Particles that Seek Stacking Faults}

\author{Bryan VanSaders}
\affiliation{Department of Materials Science and Engineering, University of Michigan, Ann Arbor, Michigan 48109, United States}

\author{Sharon C.\ Glotzer}
\affiliation{Department of Materials Science and Engineering, University of Michigan, Ann Arbor, Michigan 48109, United States}
\affiliation{Department of Chemical Engineering, University of Michigan, Ann Arbor, Michigan 48109, United States}
\affiliation{Biointerfaces Institute, University of Michigan, Ann Arbor, Michigan 48109, United States}

\date{\today}

\begin{abstract}

There is growing interest in functional, adaptive devices built from colloidal subunits of micron size or smaller.
A colloidal material with dynamic mechanical properties could facilitate such microrobotic machines.
Here we study via computer simulation how active interstitial particles in small quantities can be used to modify the bulk mechanical properties of a colloidal crystal.
Passive interstitial particles are known to pin dislocations in metals, thereby increasing resistance to plastic deformation.
We extend this tactic by employing anisotropic active interstitials that travel super-diffusively and bind strongly to stacking faults associated with partial dislocations.
We find that: 1) interstitials that are effective at reducing plasticity compromise between strong binding to stacking faults and high mobility in the crystal bulk. 2) Reorientation of active interstitials in the crystal depends upon rotational transitions between high-symmetry crystal directions. 3) The addition of certain active interstitial shapes at concentrations as low as $60$ per million host particles ($0.006\%$) can create a shear threshold for dislocation migration.

\end{abstract}

\maketitle

\section{Introduction} \label{section:introduction}

Colloidal machines assembled from sub-micron sized subunits provide a possible route to scaled down versions of so-called `particle robots' \cite{Li2019}.
Particle robots that sense and actuate are built from a collection of sub-units that, like colloidal particles, individually have limited functionality \cite{Koman2018}.
The nascent field of colloidal robotics faces many challenges, including the need for dynamically responsive colloidal materials.
Many of the functions we wish to realize are analogues to the capabilities of living cells.
Unicellular organisms accomplish functional shape changes by stiffening and softening their cytoskeletal matrix dynamically \cite{Fletcher2010}.
A prototypical example is the unicellular predator \textit{Amoeba Proteus}, which grows and reshapes extensions of itself to move and hunt.
These changes are enabled by dynamic mechanical property changes and flows within the cell \cite{Rogers2008}.
In mimicry of this functionality, researchers have investigated how the mechanical behavior of sub-$\mu m$ particle assemblies can be coupled to external fields for controllable functional changes, for example by creating stiffness changing colloidal crystals \cite{Kim2016} and magnetic `microbots' \cite{Yu2018,Yigit2019,Xie2019}.

When considering the deformation of dense arrangements of colloids, we can look to the extensive knowledge base concerning the deformation of metals to guide us.
Materials scientists have gleaned deep insight into the mechanisms of deformation for metals, particularly the importance of dislocations in plastic deformation \cite{Hirth1982}.
Under shear stress, dislocations migrate and effectively transport material though a crystal, resulting in shape change of the material at state points below the melting transition, and stresses below the ultimate yield strength.
In this regard, colloidal crystals behave similarly to metals \cite{Schall2004, Schall2006, Lin2016, vdMeer2017, VanSaders2018}.
A classical approach to impeding the motion of dislocations is to introduce impurity particles into the host crystal.
Carbon steels are the prototypical example of this tactic; carbon interstitials significantly increase the shear stress required to drive a dislocation to glide when added to iron at concentrations of $<1\%$.

Such interstitial pinning can be applied to colloidal systems to help control material deformation.
Furthermore, colloidal particles and interstitials may have many designable and exotic properties, such as anisotropic shape or self-propulsion (i.e. active matter) \cite{Glotzer2007, Sacanna2011, Nykypanchuk2008, Sacanna2011, Boles2016, Marchetti2013}.
Active interstitials allow us to consider scenarios where the solute species is capable of self-propulsion and so travels super-diffusively.
If such active interstitials can also be designed to bind strongly to dislocations and prevent their motion, then we could expect a comparatively small number of active interstitials to have a large effect on colloidal crystal plasticity.
The migration of passive interstitials to dislocations is controlled entirely by the diffusive properties of the interstitial.
As such, long timescales or high temperatures are required for passive solutes to accumulate around dislocations \cite{Cottrell1949, Yoshinaga1971, Hirth1982, Fan2013}.
Active particles could decouple the timescale of diffusion from that of dislocation pinning, as well as open the possibility of a metamaterial with dynamic, controllable plasticity by toggling activity.

In this study we explore the ways in which active anisotropic interstitials can interact with a dense crystalline environment via computer simulation.
We show that anisotropic active interstitial particles can have strong effective attractive interactions with stacking faults in crystals of spheres interacting via isotropic steep repulsive potentials.
These active particles adsorb onto the stacking faults that link dissociated dislocation pairs and thereby pin dislocations.
We show that interstitial anisotropy and active force magnitude also affect the characteristics of the path explored by the interstitial.
With the reduction of dislocation mobility as the primary goal of this study, we frame the design of active interstitials as a competition between the ability of the interstitial to move freely through the crystal bulk (so as to locate stacking fault binding sites) and the affinity of the interstitial to remain tightly bound to these sites.
We propose a combined metric that assesses the overall effectiveness of an active interstitial as a dislocation-pinning additive, and test a high-performing interstitial in a dislocation-containing system under bulk shear.
We find that for well designed active interstitials, a number concentration as low as $64$ per million host particles is sufficient to introduce significant barriers to dislocation glide and material plasticity.

\section{Model and Methods} \label{section:methods}

\subsection{The Active Interstitial Model}

\begin{figure}[t]
\centering
\includegraphics[width=\columnwidth,keepaspectratio]{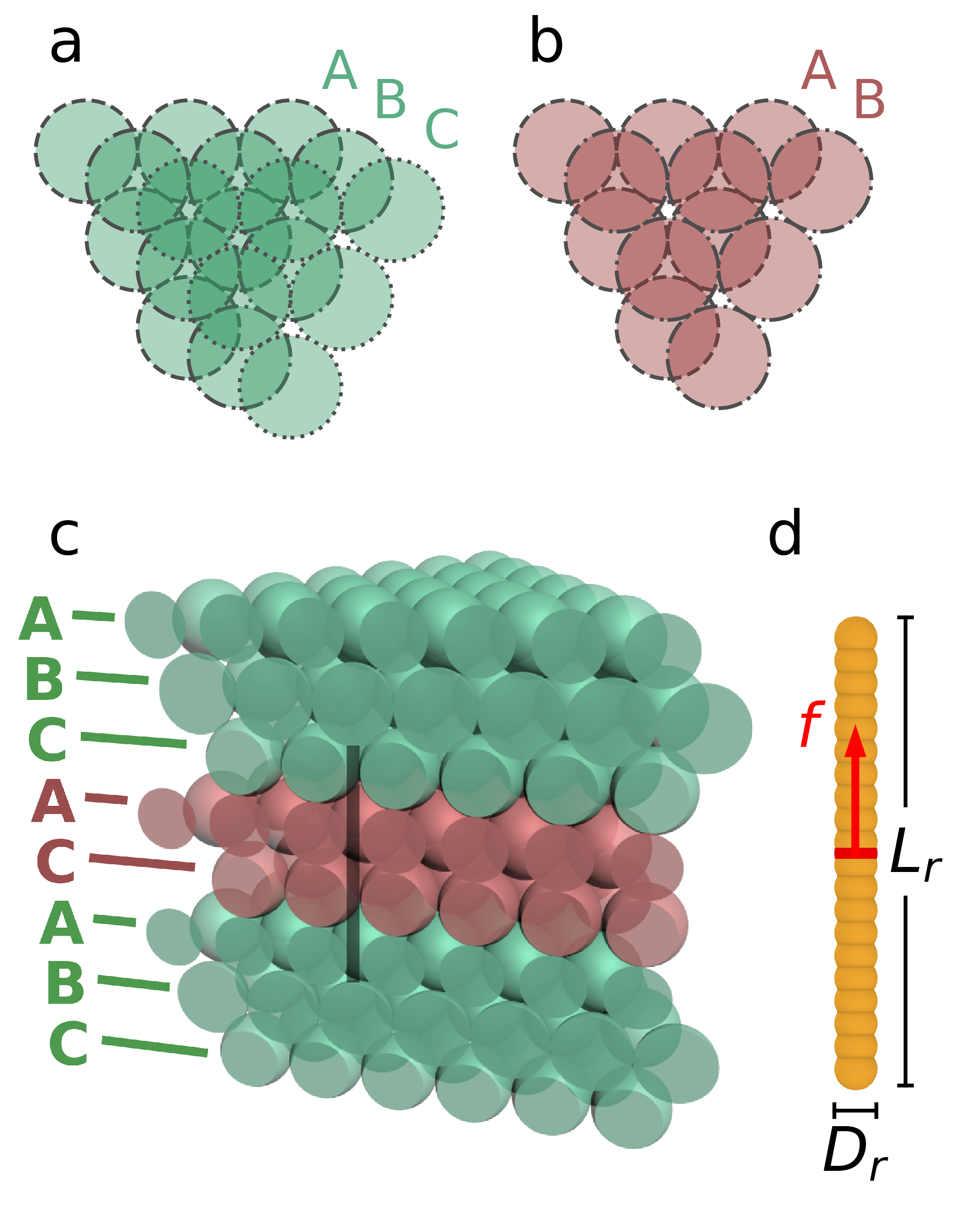}
\caption{
    \textbf{a} Stacking sequence of dense planes in the FCC crystal.
    \textbf{a} Plane stacking sequence for the HCP crystal.
    \textbf{c} Cut-away rendering of a stacking fault with cylindrical void indicated by a shaded rectangle.
    \textbf{d} A rendering of the active interstitial geometry used in this study.
    The center of mass of the interstitial is indicated with a red line.
}
\label{fig:rod_scheme}
\end{figure}

The particles that comprise the host crystal are represented by steeply repulsive isotropic pair potentials.
All particles interact via the shifted-Weeks-Chandler-Andersen potential (SWCA) \cite{Weeks1971}.
The value of $\sigma$ used in this potential was set to $0.2$, and for the host particles the potential was shifted in radius so that the minimum (located at $\sigma 2^{1/6}$ in the unshifted case) is kept at a distance of $2^{1/6}$.
This distance is hereafter referred to as $D$ (the host particle diameter).
The lowest free energy crystalline phase for such particles is face centered cubic (FCC), with hexagonal close packed (HCP) a close second \cite{Pronk1999}.
FCC and HCP can both be constructed by alternating sequences of close-packed planes of particles (Fig.~\ref{fig:rod_scheme}a-b).
Dislocations in FCC crystals can dissociate into pairs of partial dislocations that conserve total Burgers vector \cite{Hirth1982}.
These dissociated pairs are connected by a stacking fault, which is a local change of stacking sequence from FCC to HCP.
The proclivity of dislocations to dissociate is controlled by the stacking fault energy, which is the free energy penalty of stacking fault creation \cite{Hirth1982}.
In crystals composed particles interacting via short range repulsive potentials, this free energy penalty is very low, and so dislocations dissociate readily \cite{Pronk1999}.

Stacking faults are low free-energy defects for steep repulsive solids because the close-packed FCC and HCP stacking sequence motifs have no density difference.
However the topology of the void space in HCP is different than in FCC.
Fig.~\ref{fig:rod_scheme}a-b shows the stacking sequence of FCC and HCP planes, respectively.
The HCP crystal has linear, connected voids that are parallel to the densest-packed plane normals (the HCP $[0001]$ and FCC $[111]$ crystal directions).
In FCC, these voids can extend only for three layers before they are interrupted.
Fig.~\ref{fig:rod_scheme}c shows a stacking fault embedded in an FCC crystal.
This stacking fault introduces linear channels that span five stacking planes.
In a close-packed crystal of diameter $D$ spheres, this void channel has a length of $(\frac{5\sqrt{6}}{3}-1)D \approx 3.1\,D$ and minimum open diameter of $(\frac{2}{\sqrt{3}}-1)D \approx 0.15\,D$.

In this study our interstitial particle is a rod composed of spheres interacting via repulsive potentials.
We showed recently that active rod-like interstitials fit within void channels of the FCC crystal structure and have geometry-dependent mobility \cite{VanSaders2019}.
When active, the rods have three parameters (Fig.~\ref{fig:rod_scheme}d) that govern their behavior: end-to-end length ($L_r$), diameter ($D_r$), and active force magnitude ($F$).
Here activity is represented by a constant force, directed parallel to the long axis of each interstitial.
Fig.~\ref{fig:rod_scheme}d shows a schematic of the rod geometry used in this study.
We considered variable rod diameter, length, and driving force.
Forces ranging from $3 \, kT/D$ to $45 \, kT/D$ were applied to rods with maximum distance between beads $0.5D$ to $3D$ and diameters $0.2D$ to $0.4D$.
Total end-to-end rod length ($L_r$) is equal to the distance between the centroids of the end beads plus the bead diameter.

\subsection{Molecular Dynamics Methods} \label{subsection:md_meth}

All molecular dynamics (MD) simulations reported here were performed with \texttt{HOOMD-blue} (v2.0) \cite{Anderson2008, Glaser2015}.
Simulations were performed in the $NPT$ ensemble as derived by Martyna et al.\ \cite{Martyna1996}.
System thermal energy was held at $kT=0.1$, pressure at $P=2$, and host particle mass was fixed at $m=10$ (in simulation units).
At this statepoint the SWCA potential results in behavior similar to hard spheres \cite{Filion2011,VanSaders2018}.

In all cases discussed, the interstitial was simulated as a collection ($N=20$) of isotropic repulsive potentials (interacting via the same SWCA potential as the host particles).
To produce rods of different diameter the radial shifting of the SWCA potential was varied.
For the rod-like interstitial relative positions were maintained and torques handled during MD integration by rigid-body simulation \cite{Nguyen2011}.
The mass and thus moment of inertia of the interstitial was set so that the mass density of the volumes enclosed by the zero isoenergy surfaces of the host and interstitial particles are equal.

\section{Sampling Protocols}

\subsection{Interstitial Binding Protocol} \label{subsection:binding_sampling}

To explore the interaction of active anisotropic particles with void spaces present near stacking faults in a thermalized, non-close packed crystal of SWCA particles, we prepared periodic simulation domains ($N=20,480$) with single stacking faults (using a layer shifting pattern of $3\cdot[ABC]+ BC + 3\cdot[ABC]$).
A single rod particle with axial force was added.
By observing the position of the active particle relative to the stacking fault, a one-dimensional (the $[111]$ crystal direction) probability density function was sampled.
In cases of strong particle-void interaction, we observed a sharp peak in probability density near the stacking fault.

\subsection{Active Walk Protocol} \label{subsection:walk_sampling}

To investigate the mobility of active interstitials, we conducted MD simulations of individual active interstitials in defect-free crystalline domains with periodic boundaries ($N=18,432$).
The volume swept out by the trajectory of an active walk was computed using the open source software \texttt{vorlume} contained in the Structural Bioinformatics Library package \cite{Cazals2017}.
The path of the active particle was decorated with spheres as the input to \texttt{vorlume}.
Sphere centers were placed on the path at separations no larger than one eight of their diameter (interpolated as needed) to approximate a cylindrical swept volume.

\subsection{Dislocation Interaction with Active Interstitial Protocol} \label{subsection:shear_sampling}

Dislocation line arrays were created by subtracting a half plane of particles in a periodic simulation box ($N=502,500$) spanning 60 unit cell lengths in the $x$ direction (aligned with crystal direction $[1 \bar 1 0]$), 15 unit cell lengths in the $y$ direction (aligned with crystal direction $[1 1 \bar2]$), and 46 unit cell lengths in the $z$ direction (aligned with crystal direction $[111]$).
Active particles were introduced with directors perpendicular to dislocation glide planes.
After sufficient time to allow the active particles to explore the simulation domain, a shear stress ($\sigma_{xz}$) was applied to the crystal, driving the dislocations to glide.
System shear strain was computed from the simulation's box matrix \cite{Parrinello1982}.

\section{Results and Discussion} \label{section:results}

\subsection{Effective Interstitial Attraction to Stacking Faults} \label{subsection:binding}

\begin{figure}[t]
\centering
\includegraphics[width=\columnwidth,keepaspectratio]{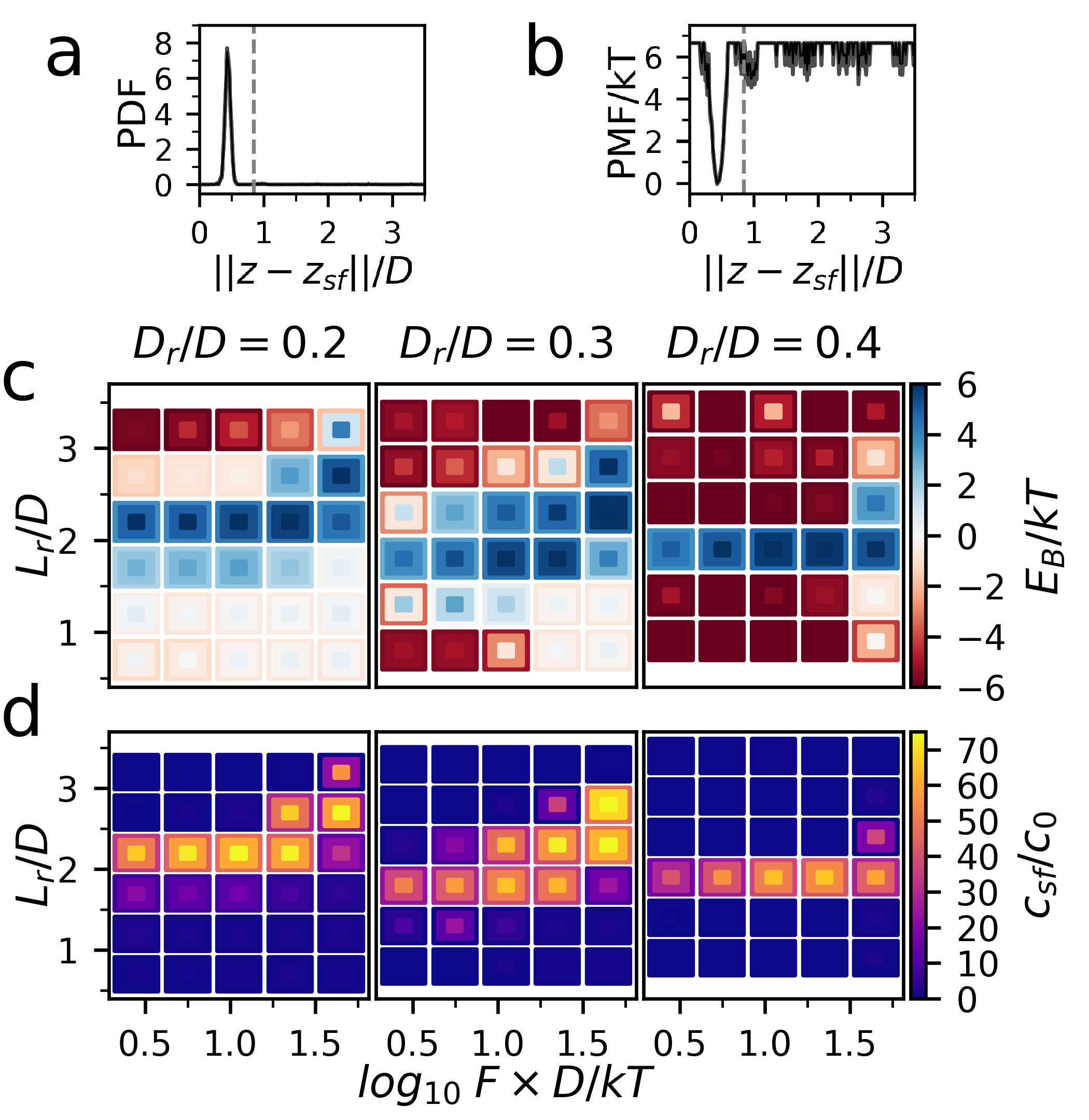}
\caption{
    \textbf{a} A typical probability distribution of the active interstitial's position relative to the stacking fault center.
    The dashed line indicates the cutoff used to define interstitial-stacking fault contact.
    \textbf{b} Estimate of interaction free energy obtained from sampled probability distribution.
    \textbf{c} Binding free energy well depth for interstitials of various parameters.
    Each pixel represents both mean value and error: the centermost ring is the mean value of localization, the middle and outer rings are +/- one standard deviation.
    Force is non-dimensionalized by host particle diameter ($D$) and system thermal energy ($kT$).
    \textbf{d} Concentration enhancement for interstitials of different parameters.
}
    \label{fig:sf_binding}
\end{figure}

We find that active interstitials have a geometry-dependent effective attraction to stacking faults.
To quantify this emergent attraction, we estimate a probability density function from observations of the interstitial stacking fault separation (Fig.~\ref{fig:sf_binding}a shows an example of such a distribution).
From this pair correlation function we obtain a potential of mean force \cite{Chandler1985}, yielding an estimate of the interaction free energy of the interstitial with the stacking fault.
Fig.~\ref{fig:sf_binding}b shows an example of such a free energy curve.
Fig.~\ref{fig:sf_binding}c shows the estimated binding free energy well depth ($E_B$) for all rod geometries and driving forces explored in this study.
Negative values of $E_B$ indicate that the interstitial was found near the stacking fault less frequently than elsewhere in the simulation domain.
We can also estimate the concentration of interstitials expected to accumulate near a stacking fault.
We estimate the concentration enhancement ($c_{SF}/c_{0}$) from the probability per volume of observing the rod within a cutoff from the stacking fault (dashed line, Fig.~\ref{fig:sf_binding}a and b), normalized by the total volume of the simulation (Fig.~\ref{fig:sf_binding}d).

We find that the length of the interstitial is the critical parameter controlling the strength of effective attraction; interstitials with $L_r \approx 2\,D$ have the deepest free energy binding wells and highest concentration enhancements.
Active force magnitude and interstitial diameter have significantly less effect on binding characteristics.
Interstitial lengths longer than $2\,D$ tend not to be effective at binding to stacking fault voids.
This is in part due to the slow rotational dynamics of long rods in crowded environments; some long rod geometries failed to align with the voids present near the stacking fault.

\subsection{Considering the Path of an Active Interstitial as a Search}

Beyond binding, we can ask the question, `is the active walk of an interstitial an efficient search pattern for the target site (i.e. stacking fault) of interest?'
To address this question, we first analyze the statistics of the active interstitial's trajectory through the host crystal in the absence of defects.
The question of which form of run-and-tumble active walk is most effective for locating a target has been addressed at length, particularly in reference to the search patterns of animals \cite{Viswanathan1999,Bartumeus2003,Reynolds2009,Humphries2012}, but also in relation to general classes of search problems \cite{Hakli2014}.
Rupprecht et al.~analyzed the mean first passage time in a model system and found that the statistics of run lengths affects searching efficiency \cite{Rupprecht2016}.
They found that the optimal run length depends upon the distribution of targets, with dense target environments favoring a Gaussian distribution and sparse environments favoring long-tail distributions.
Furthermore, optimal run length also depends upon the boundary conditions of the domain considered, and the domain size.

The distribution of stacking faults due to dissociated dislocations will depend upon the size of the crystal domain in question, since dislocations have long-range strain field interactions.
Generally, experiments of colloidal crystals do not exceed centimeters in scale ($\approx 10^4$ to $\approx 10^5$ sub-$\mu m$  particles).
In dislocation terms, this is a small domain, and so a small number of dislocations should be expected in samples of reasonable quality.
We expect, therefore, that active walks with long-tailed distributions of run lengths should be the best performers.
When considering experimental systems we usually imagine a colloidal crystal domain with open boundaries, in which case searchers that encounter the boundary will be lost (either by leaving the crystalline domain, or becoming trapped on the boundary).
Therefore we desire active walks that have a high probability of encountering a stacking fault before they are lost from the system boundaries.
Another way to state this goal is to say we want a space filling walk; a walk that travels long distances slowly but samples large volumes without repetition.

\begin{figure}[t]
\centering
\includegraphics[width=\columnwidth,keepaspectratio]{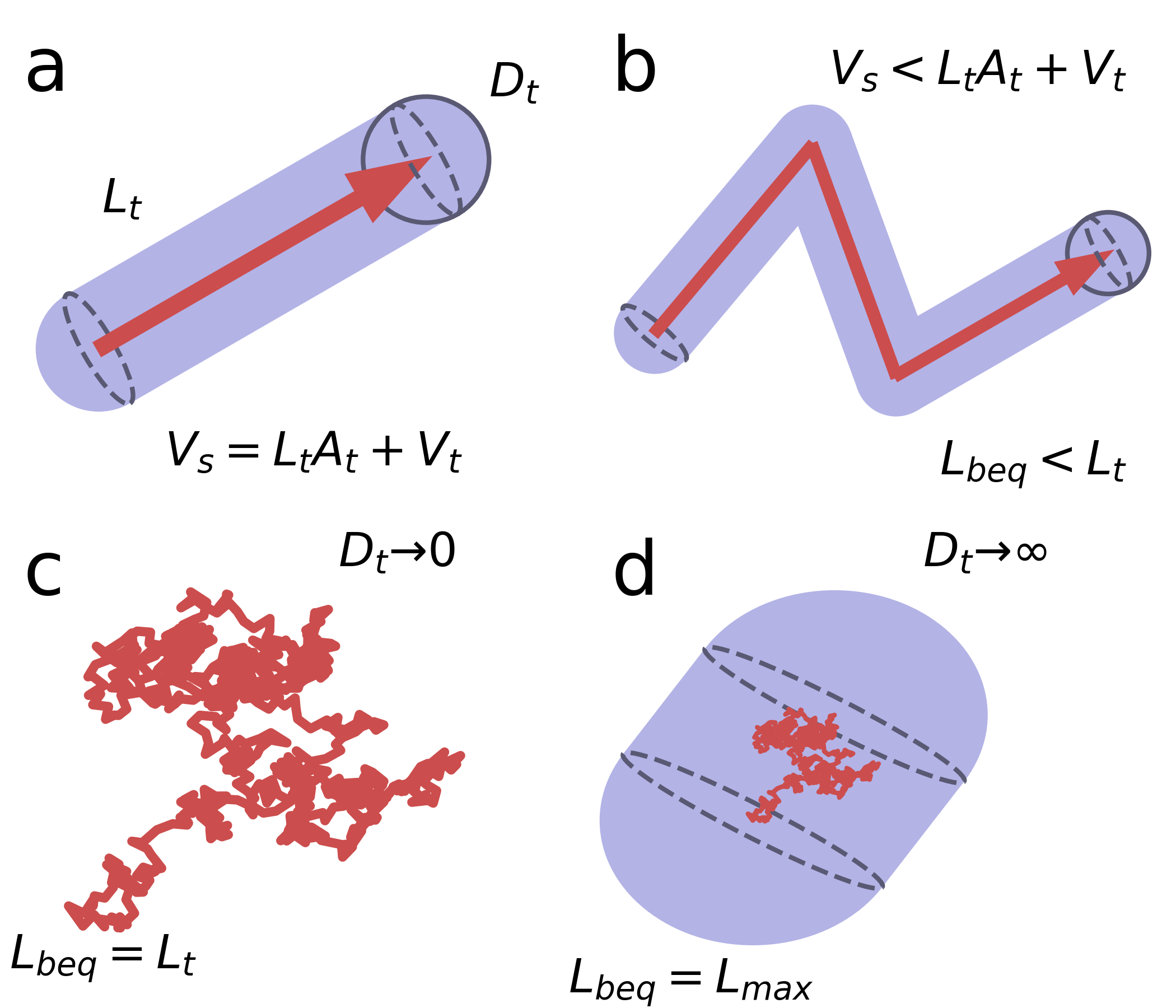}
\caption{
    \textbf{a} The volume (blue) swept out by a ballistic walk (red).
    \textbf{b} The volume swept out by a run-and-tumble walk.
    \textbf{c} A walk with many steps decorated with infinitesimal $D_t$.
    \textbf{d} A walk with many steps decorated with very large $D_t$.
}
    \label{fig:Lbeq_schematic}
\end{figure}

We envision the active walk of an interstitial as a swept volume, where the sweeping diameter ($D_t$) is twice the walker-target collision distance.
In this study we treat the walker as a point object, and consider the effect of target size.
Consider a purely ballistic walk (Fig.~\ref{fig:Lbeq_schematic}a).
In such a walk, the swept volume is simply related to the total length of the walk ($L_t$) and the cross-sectional area ($A_t$) and volume ($V_t$) of sweeping.
The total swept volume of the spherical interaction area can be described as $V_s = L_t A_t + V_t$.
We refer to a walk described by this equation as `ballistic'.
Were we to calculate the ballistic length ($L_{beq}$) of a walk that was not actually ballistic, we would find that this equivalent length is less than the total length of the interstitial's trajectory (Fig.~\ref{fig:Lbeq_schematic}b).
The degree to which $L_{beq} \neq L_t$ depends upon the diameter of the walk.
For infinitesimal values of walk diameter, $L_{beq} = L_t$ regardless of the number of bends in the walk (Fig.~\ref{fig:Lbeq_schematic}c); for very large values of $D_t$, $L_{beq}$ will be equal to the largest distance of the walk (Fig.~\ref{fig:Lbeq_schematic}d).
For simulated walks, we generally find that $L_{beq}$ as a function of $D_t$ begins at a high value, and transitions to a low value at a specific $D^*$ that depends upon the geometry of the walk.
Consequently, the quantity $-\partial L_{beq} / \partial D_t$ is singly peaked.
The location and height of this peak supplies information about the space-filling properties of the walk geometry.
Significantly below $D^*$, the active walk has few self overlaps, and so the space sampled by the interstitial is almost entirely new.
Above $D^*$, there are significant overlaps between segments, and therefore significant portions of the space are sampled multiple times.
When $D_t = D^*$, the walk is most nearly space-filling.
The peak magnitude indicates how nearly space-filling the walk is.
A large derivative indicates that large lengths of path are at a distance $D^*$ from each other.
These geometric measures of space-filling performance are also functions of time.
After a sufficiently long time all walks (of finite-sized searchers) that are not strictly ballistic will become diffusive, since the walker will eventually re-sample previously visited space.
We propose that for desirable active walks the peak in $-\partial L_{beq} / \partial D_t$ should be centered on the target size of interest, and the roughness should evolve slowly in time so that the walk remains an efficient space-filling search over the timescale of interest.
Therefore, when evaluating active walks we consider the time averaged value of $-\partial L_{beq} / \partial D_t$, denoted as $\langle-\partial L_{beq}/ \partial D_t\rangle_t$.

\subsection{$L_{beq}$ as a Function of Interstitial Parameters}

\begin{figure}[t]
\centering
\includegraphics[width=\columnwidth,keepaspectratio]{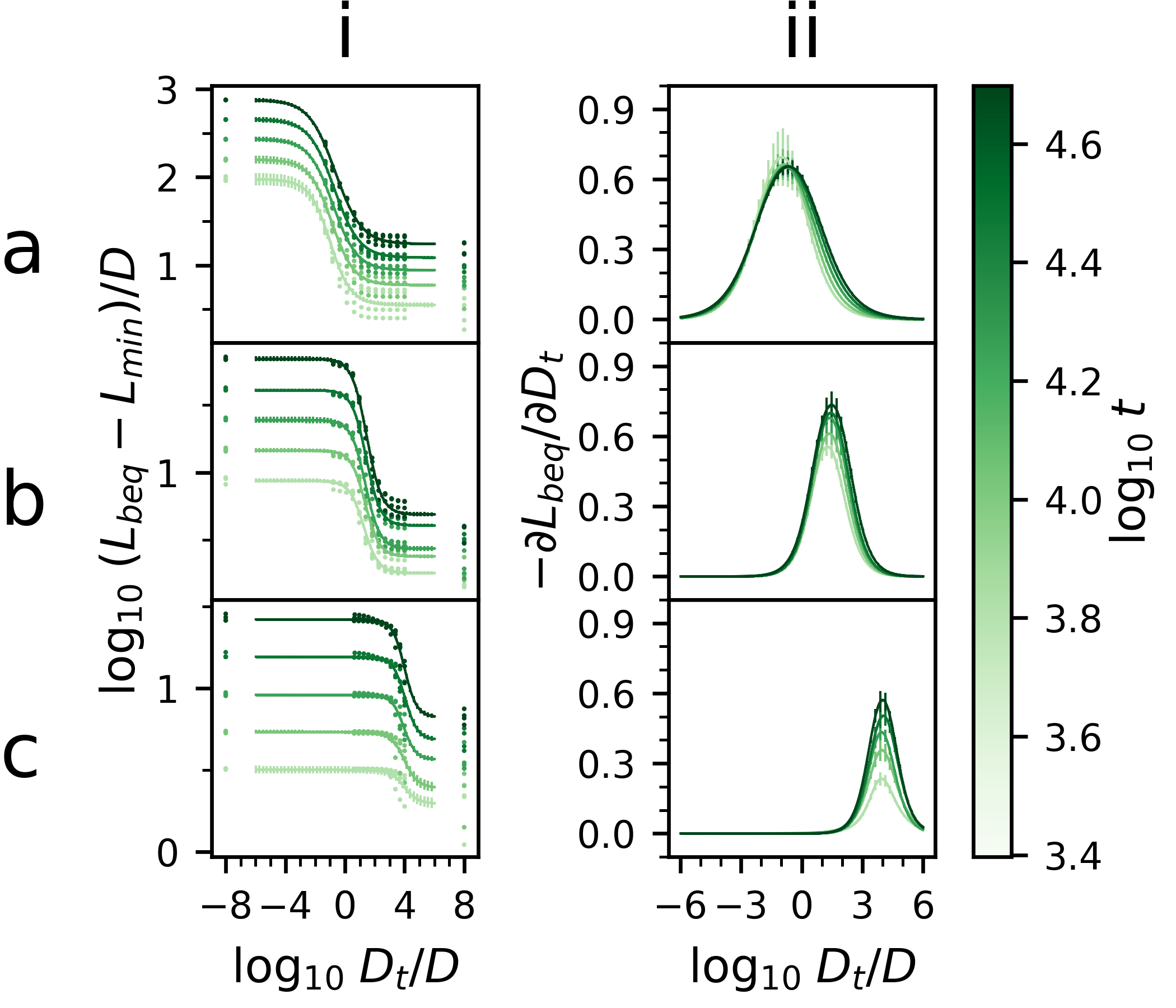}
\caption{
    $L_{beq}$ (\textbf{column i}) and $-\partial L_{beq}/\partial D_t$ (\textbf{column ii}) for example active walks.
    \textbf{row a} A diffusive walk.
    \textbf{row b} A rough walk.
    \textbf{row c} A ballistic walk.
}
    \label{fig:Lbeq_examples}
\end{figure}

We investigated the space-filling properties of active interstitial walks by analyzing the $L_{beq}$ value of interstitials with various $[F,L_r,D_r]$ parameters.
Fig.~\ref{fig:Lbeq_examples} shows three examples of active walks.
Column (i) shows $L_{beq}$ for diffusive (a), rough (b), and ballistic (c) walks.
Shown data points are sampled from four replicates of each walk.
Asymptotic values (shown here at $\log_{10} D_t/D=-8$ and $+8$) are calculated from the total path length and largest distance in the trajectory point cloud, respectively.
We fit the data to a logistic function of the form:

\begin{equation}
    \log_{10}\, L_{beq} = \frac{\log_{10}\,L_i - \log_{10}\,L_f}{1-e^{k(\log_{10}\,D_t-\log_{10}\,D^*)}} + \log_{10}\,L_f
\end{equation}

\noindent where $L_i$ is the total path length, $L_f$ is the largest point cloud distance, $D^*$ is the transition target diameter, and $k$ is the steepness of the transition.
Fig~\ref{fig:Lbeq_examples} column (ii) shows the value $-\partial L_{beq} /\partial D_t$.
We find that the transition from diffusive to ballistic is clearly captured in the location of the $L_{beq}$ transition point.
As an active interstitial walk becomes more ballistic, the target diameter for which the walk most nearly fills space increases.
Ballistic (diffusive) walks are best for very large (small) target sizes.
Rough walks are intermediate between the two.

\begin{figure}[t]
\centering
\includegraphics[width=\columnwidth,keepaspectratio]{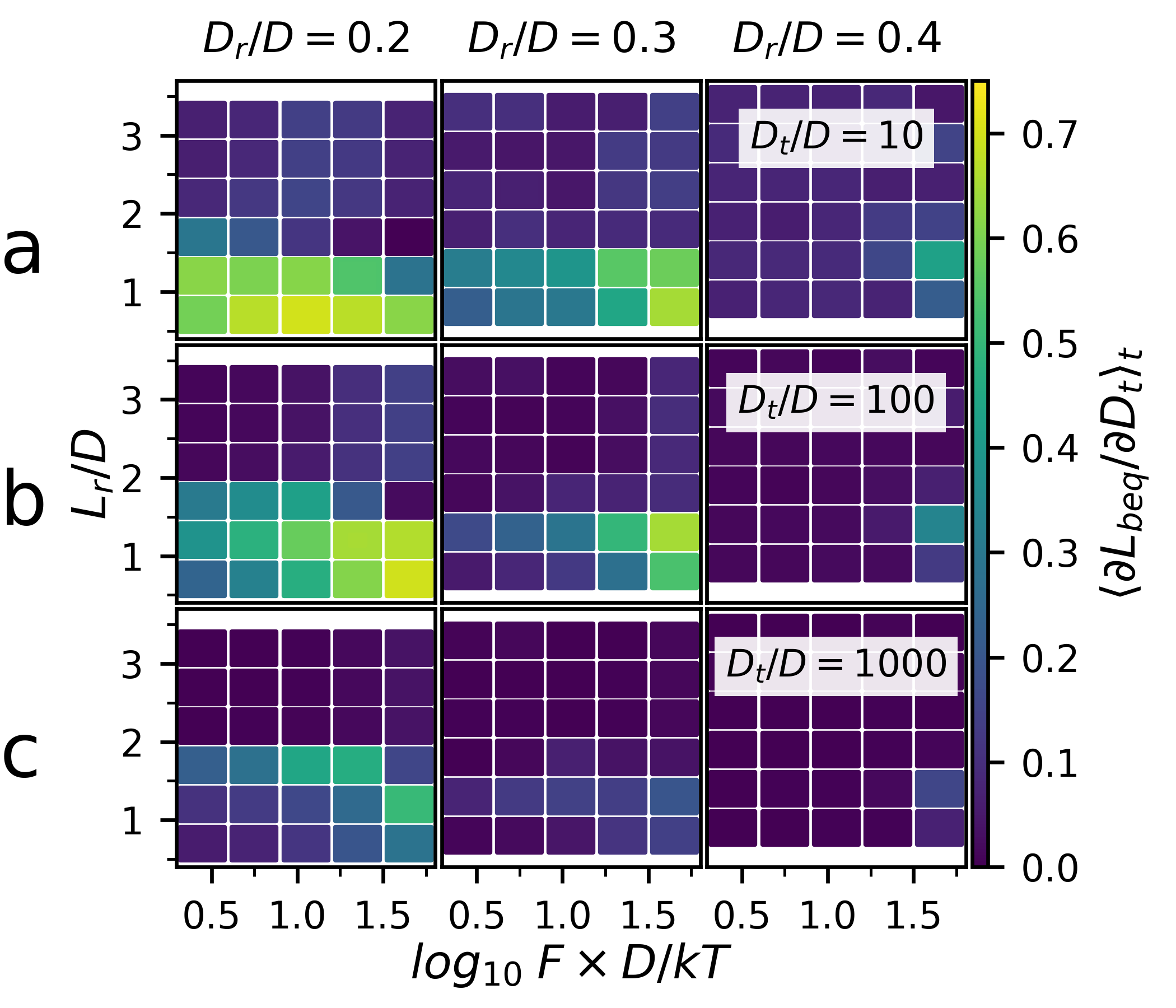}
\caption{
    Time averaged derivative of $L_{beq}$ evaluated for different values of $D_t$ for all active interstitial parameters.
    \textbf{a} $D_t=10$.
    \textbf{b} $D_t=100$.
    \textbf{c} $D_t=1000$.
    Error is represented the same as in Fig.~\ref{fig:sf_binding}, however these errors are small.
}
    \label{fig:dLdD_trends}
\end{figure}

Fig.~\ref{fig:dLdD_trends} shows the value $-\langle \partial L_{beq} /\partial D_t \rangle_t$ evaluated at different values of $D_t$.
We find that walks are most capable of filling space for interstitial lengths less than $2D$.
This finding is in direct opposition to the trends of strong binding to stacking faults reported in section \ref{subsection:binding}, where the largest effective stacking fault-interstitial attractions were found to be for rods of length $2D$.
The ability of an active interstitial to search space depends upon a low barrier to reorientation; however, a low barrier to reorientation also permits the interstitial to escape from the void present near the stacking fault.
As a result, the binding and mobility of active interstitials must be balanced against each other in a compromise between these two effects.

\subsection{Reorientation Probability Controls Walk Roughness}

\begin{figure}[t]
\centering
\includegraphics[width=\columnwidth,keepaspectratio]{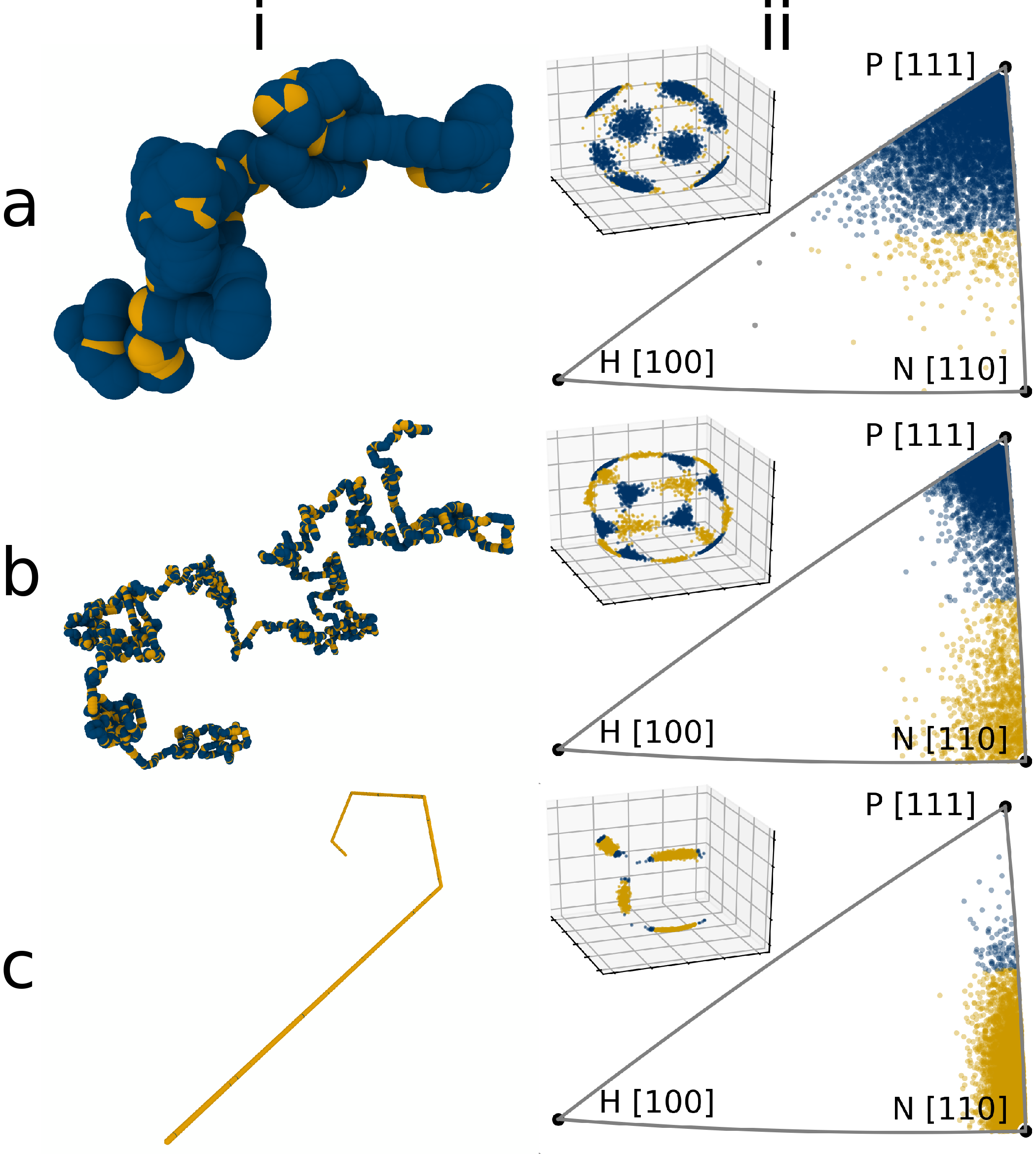}
\caption{
    \textbf{column i} Renderings of volume swept by active walks (using \texttt{Ovito} \cite{OVITO}).
    Points along the walk are colored by the high symmetry crystal direction that the active interstitial's director is most closely pointing towards.
    \textbf{column ii} Directors of the interstitial as it performs the active walk, mapped onto the irreducible symmetry zone of the FCC structure.
    Insets show the un-reduced cloud of directors.
    \textbf{row a} A diffusive walk.
    \textbf{row b} A rough walk.
    \textbf{row c} A ballistic walk.
}
    \label{fig:orientation}
\end{figure}

Beyond the path traced by the active interstitial rod during the  walk, (Fig.~\ref{fig:orientation}(i)), we can also investigate the orientation of the active interstitial relative to high-symmetry crystal directions.
Fig.~\ref{fig:orientation}a(i) shows the volume swept out by an active interstitial that migrates diffusively.
Points along the walk are colored by the nearest high symmetry crystal direction towards which the interstitial's long axis was pointed.
Fig.~\ref{fig:orientation}a(ii) shows a polar projection of these directors mapped onto the irreducible rotational zone of the FCC crystal.
The inset shows the un-reduced director cloud.
Fig.~\ref{fig:orientation}c(i)-c(ii) and d(i)-d(ii) show the same information for a `rough' and ballistic walk, respectively.

We find that diffusive, rough, and ballistic walks all have distinct orientational signatures.
Interstitials with diffusive behavior tend to remain aligned with the $[111]$ family of directions (which we denote as $P_{[111]}$).
In the close-packed FCC structure, cylindrical voids of length $(\sqrt{6}-1)D \approx 1.45D$ aligned with $P_{[111]}$ exist.
Diffusive interstitials are trapped within these isolated voids, and so are unable to cover significant distances.
In contrast, ballistic walks tend to remain aligned with the $N_{[110]}$ family of directions.
Close packed sphere FCC crystals have small diameter channels aligned with these directions that extend indefinitely.
Ballistic interstitials have a high barrier to reorientation, and we find that $N_{[110]}$ channels are preferred over $P_{[111]}$ channels, but transitioning between members of the $N_{[110]}$ family is difficult.
Rough walks occupy both families of directions.
These interstitials are able to explore many orientations because there is a relatively small barrier to switching between the $P_{[111]}$ and $N_{[110]}$ directions.

\begin{figure}[t]
\centering
\includegraphics[width=\columnwidth,keepaspectratio]{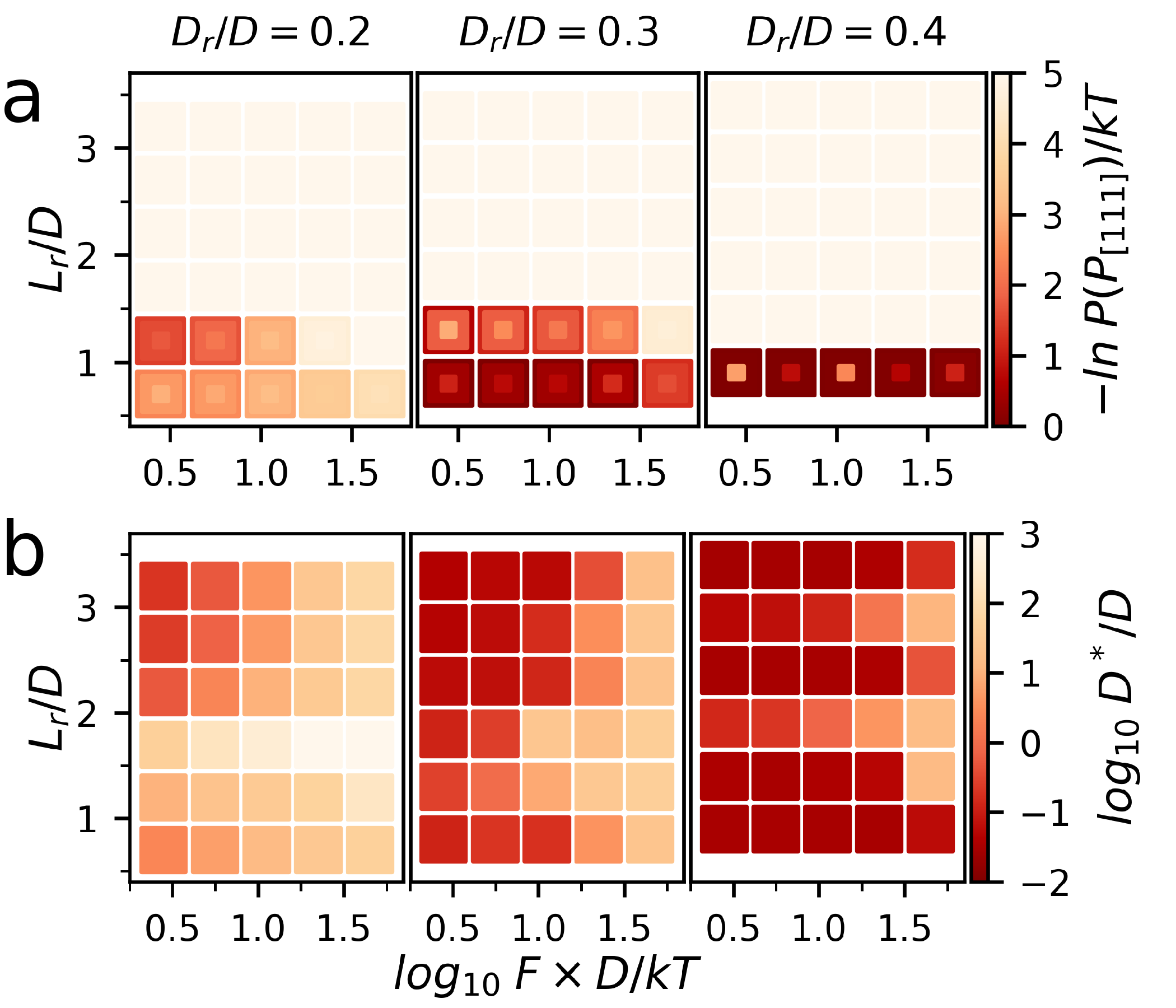}
\caption{
    \textbf{a} Estimate of the free energy of the $P_{[111]}$ state for interstitials in this study.
    Error is represented as in Fig.~\ref{fig:sf_binding}.
    \textbf{b} The best-fit value of the inflection point of the transition from ballistic to diffusive behavior ($D^*$).
}
    \label{fig:discussion}
\end{figure}

Ultimately, we find that the ability of a rod-like active interstitial to reorient controls both the binding to stacking fault voids and the mobility characteristics.
The ability of the interstitial to rotate can be shown to be primarily dependent upon the end-to-end length of the particle.
Fig.~\ref{fig:discussion}a shows the free energy of interstitials aligned with the $P_{[111]}$ direction in a defect-free crystal, as estimated by the probability of observing that state.
There is a clear transition from low to high energy at $L_r \approx 1.5\,D$.
The $P_{[111]}$ family of directions in an FCC crystal coincide with cylindrical voids of length $\sqrt{3}a-D$, with $a$ the lattice constant.
For the state point studied here, $a \approx 1.04\sqrt{2}D$, and so these voids have average length $~1.55\,D$.
Figure \ref{fig:discussion}a shows interstitials of this length have a free energy penalty of $2-3\, kT$ when occupying these voids.
Comparing this to Figure \ref{fig:sf_binding}c and d, we can see that strong binding and interstitial concentration occur at lengths slightly larger than this.
This suggests that strong confinement in the stacking fault voids, which are of length $5\sqrt{3}a/3-D \approx 3.25\,D$, occurs when the P family channels in the bulk are unfavorable enough to be rarely observed, but not so unfavorable as to be totally inaccessible.

By plotting the best fit value of $D^*$ (Fig.~\ref{fig:discussion}b) for the rod geometries studied here we can see that the length scale of active walk roughness is maximum for rod lengths just longer than the transition from low to high $P_{[111]}$ free energy ($L_r > 1.5\,D$).
The $P_{[111]}$ channels that are present in FCC are wider in average diameter than the $N_{[110]}$ channels ($\sim0.15\,D$ and $\sim0.07\,D$, respectively).
Therefore when the total interstitial length is short enough to fit within the $P_{[111]}$ channels, they are significantly better sites than the $N_{[110]}$ channels.
However these channels are of limited length, and so interstitials strongly bound to them tend to become caged.
Interstitials significantly longer than voids aligned with $P_{[111]}$ directions are trapped entirely in the $N_{[110]}$ channels, and are therefore able to travel in long ballistic paths but have difficulty reorienting.
When interstitial length is matched to the $P_{[111]}$ void length these voids act as transition barriers between $N_{[110]}$ channels.
Adjustments to the length of the interstitial therefore change the energy barrier for reorientation.
This barrier sets the statistics of the active walk by controlling the probability of `tumbling' between $N_{[110]}$ channels.
These results suggest that more complex interstitial geometries engineered to control rotational dynamics may be able to outperform the simple rods studied here.
Interestingly, similar dynamics of bacterial swimmers confined in porous colloidal media have recently been observed \cite{Bhattacharjee2019}.
Our results suggest that spatially structured colloidal environments may permit greater control of bacterial motion.

\subsection{A Combined Metric to Identify Optimal Interstitial Parameters}

If we consider a ballistic interstitial searching for targets of disk-like shape, then such a particle with travel length $L_t$ will possibly encounter up to $L_t/L_0$ disks, where $L_0$ is the width of one disk.
We estimate this width as twice the distance between the stacking fault center and the interstitial binding well (dashed line, Fig.~\ref{fig:sf_binding}a).
The ballistic path swept out can be written as $V_s = A_tL_t + V_t$, $L_t = (V_s-V_t)/A_t$ (which is the same expression used for defining $L_{beq}$ when the path is not strictly ballistic).
Therefore, we estimate the total number of target disks that could be encountered as $L_{beq}/L_0$.
Absorbing onto these disks will accrue a free energy benefit of $E_B$ each.
Therefore we can say that a particular walk has a free energy `potential' of $E_t = E_B L_{beq} / L_0$.
This expression represents the maximum binding free energy that can be released over the course of a walk; it is a way to assign an energy value to the distance that a walker covers.
Since $E_B$ and $L_0$ are independent of target size, we see that $\partial E_t / \partial D_t = (\partial L_{beq}/ \partial D_t)E_B/L_0$.
We can combine the negative time average of this metric with the concentration enhancement factor (see section \ref{subsection:binding}) to yield a final overall fitness score:

\begin{equation}
S = \frac{E_B}{L_0} \frac{c}{c_0} \langle -\frac{\partial L_{beq}}{\partial D_t}\rangle_t
\end{equation}

\begin{figure}[t]
\centering
\includegraphics[width=\columnwidth,keepaspectratio]{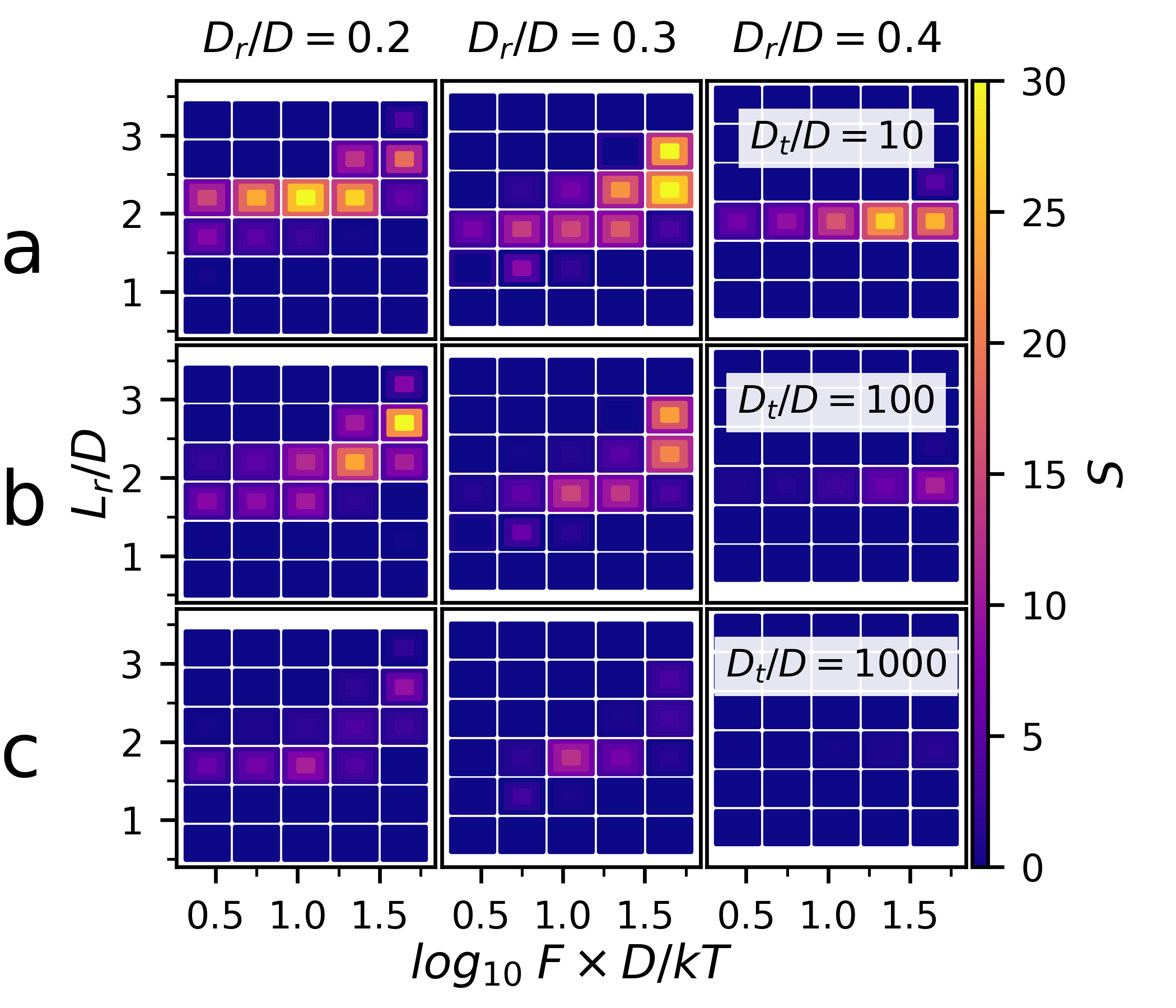}
\caption{
    Combined metric of interstitial performance ($S$) for all active interstitial parameters, evaluated at different values of $D_t$.
    \textbf{a} $D_t=10$.
    \textbf{b} $D_t=100$.
    \textbf{c} $D_t=1000$.
    Error is represented the same as in Fig.~\ref{fig:sf_binding}.
}
    \label{fig:S_trends}
\end{figure}

\noindent The result of this combined metric, evaluated at different values of $D_t$, is shown in Fig.~\ref{fig:S_trends}.
The best performing active interstitial parameters are those that are on the edge between the strongly binding (or concentrating) shapes and the highly mobile shapes.
Especially for large target sizes, a small number of optimal parameter combinations appear.

\subsection{Test Case: Pinning Dislocations to Inhibit Shear Deformation}

\begin{figure}[t]
\centering
\includegraphics[width=\columnwidth,keepaspectratio]{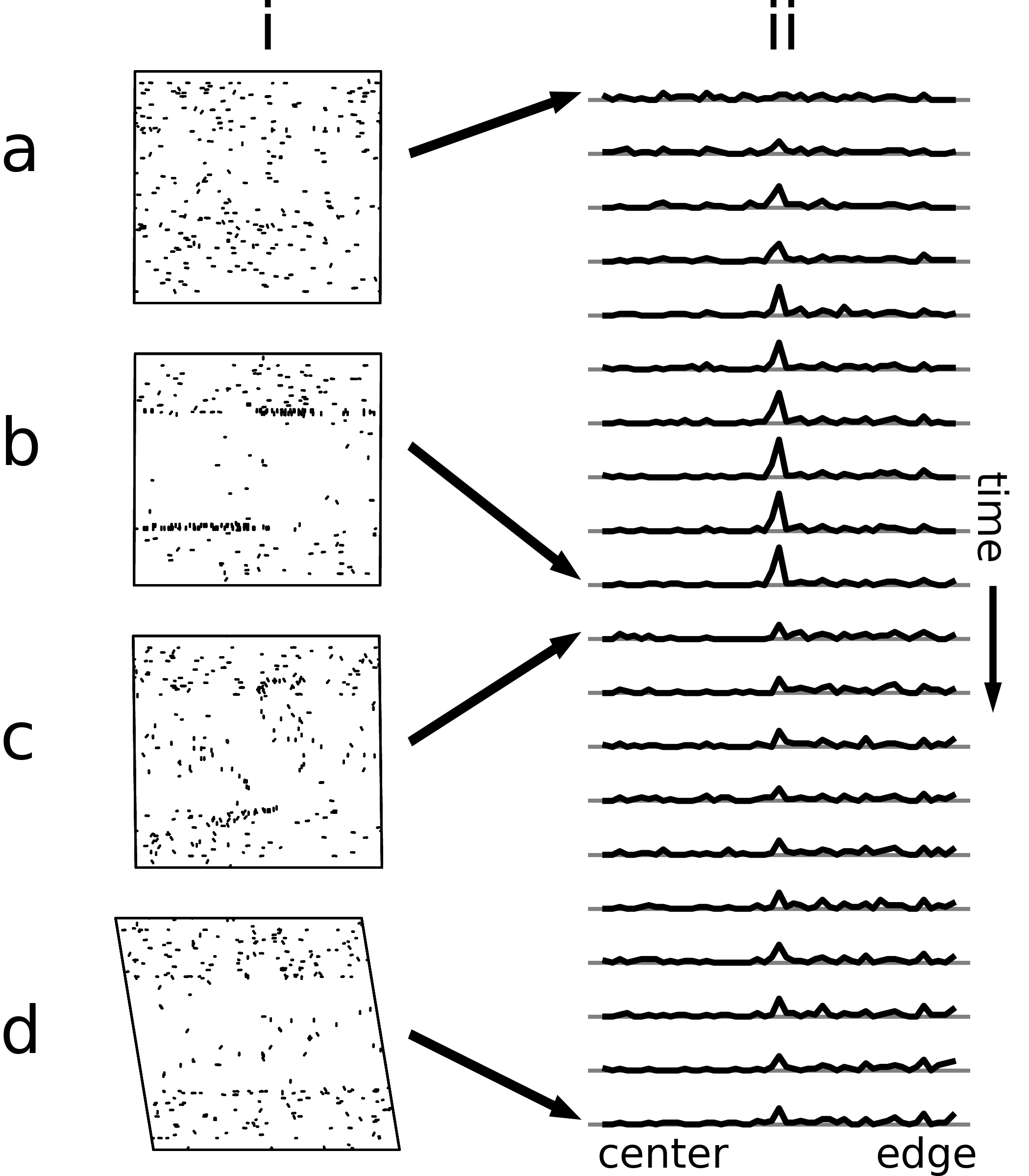}
\caption{
    Time evolution of an $N=502,500$ particle system with two dissociated edge dislocations and 256 active interstitials (with parameters $F=11.2\,D/kT$, $L_r=1.8\,D$, and $D_r=0.3\,D$).
    \textbf{column i} snapshots of the active interstitials in the simulation box, looking along the $[11\bar 2]$ crystal direction.
    Host particles are not shown.
    \textbf{column ii} Histogram of active interstitial $z$ position over the course of a shearing simulation.
    This histogram ranges from the box center to box edge.
    \textbf{a} Before interstitials have had time to encounter the partial dislocations.
    \textbf{b} Immediately before dislocation depinning. The interstitials are most concentrated in the glide planes.
    \textbf{c} De-pinning in progress. Clouds of interstitials leaving the glide plane are visible.
    \textbf{d} After several transits of depinned dislocations across the box.
    There remains a small concentration of active interstitials in the glide plane.
    Active interstitials have been depleted from the central region (under local tension) and accumulated in the edge regions (under local compression).
}
    \label{fig:absorption}
\end{figure}

We investigated the effect of stacking-fault attracted active interstitials on dislocation mobility by simulating crystalline domains with a single dislocation dipole under shear.
Using a geometry with high combined $S$ value ($F=11.2\,D/kT$, $L_r=1.8\,D$, and $D_r=0.3\,D$), active rods were introduced to the dislocation-containing crystal and allowed to walk.
Subsequently, we applied a shear force and monitored the shear strain of the simulation domain.
Fig.~\ref{fig:absorption} shows the process of dislocation pinning and shear-induced depinning.
Initially, dispersed interstitials (Fig.~\ref{fig:absorption}a(i)) search the simulation box and accumulate on the stacking faults that link the partial edge dislocations (Fig.~\ref{fig:absorption}b(i)).
At sufficient shear stress ($\sigma_{xz}$) the dislocation moves past these pinning particles, ejecting them (Fig.~\ref{fig:absorption}c(i)).
When mobile, dislocations reach a terminal glide speed that is a function of the active interstitial concentration.
Active interstitials with high $S$ values are capable of traveling rapidly enough to re-acquire slow moving dislocations, and so a persistent interstitial drag atmosphere is established (Fig.~\ref{fig:absorption}d(i)).
Fig.~\ref{fig:absorption}ii shows a histogram of active interstitial positions throughout the pinning and depinning process.
We find that active interstitials tend to accumulate into the regions under compressive strain (due to the dislocation array).
This is likely due to the larger drag experienced by active interstitials in compressed crystalline environments \cite{VanSaders2019}.

\begin{figure}[t]
\centering
\includegraphics[width=\columnwidth,keepaspectratio]{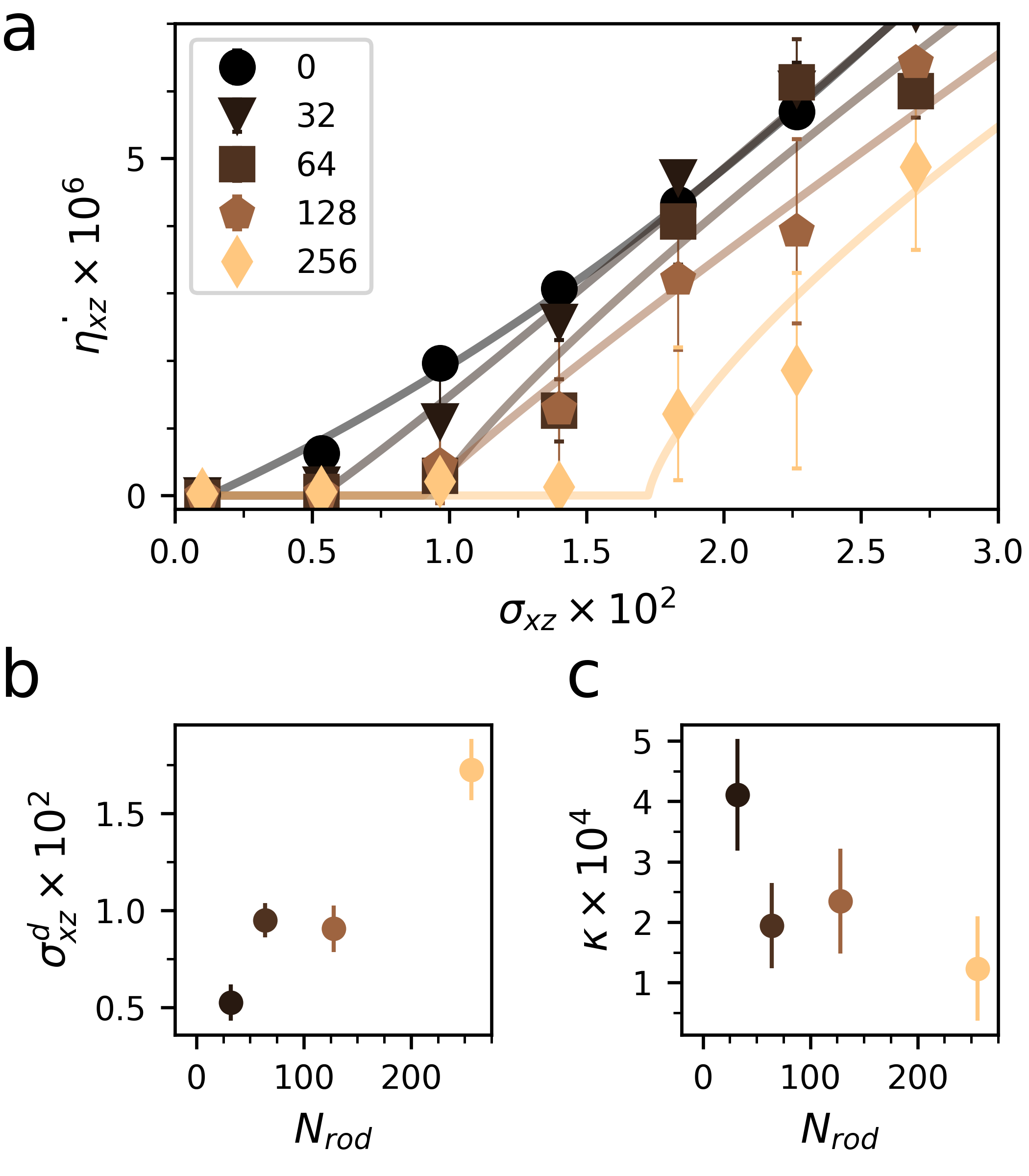}
\caption{
    Quantitative trends in shearing resistance with active interstitial number.
    Active interstitials have parameters $F=11.2\,D/kT$, $L_r=1.8\,D$, and $D_r=0.3\,D$.
    \textbf{a} Shear rate as a function of shear stress.
    \textbf{b} The depinning shear stress as a function of interstitial number.
    \textbf{c} The slope of the shear rate vs. shear stress curve as a function of interstitial number.
}
    \label{fig:shear_trends}
\end{figure}

By tracking the shear of the simulation box, the shear rate ($\dot \eta_{xz}$) can be measured for different values of $\sigma_{xz}$.
In the case of no active interstitials, the domain deforms at a rate that is best fit by a quadratic function of applied shear stress.
The intercept of this fit (Fig~\ref{fig:shear_trends}a, darkest curve) is zero to within measurement error.
A presumably small initial barrier to dislocation glide exists in such systems; however this value is not resolvable under the conditions used here.
We find that the addition of even 32 active interstitials changes this shear rate behavior from quadratic to thresholded linear.
We fit the data by curves of functional form:

\begin{equation}
\dot \eta_{xy} = max(0,\kappa |\sigma_{xy}-\sigma^d_{xy}|^\alpha)
\end{equation}

\noindent where $\sigma^d_{xy}$ is the shear threshold and $\kappa$ is the shear rate.
As the number of interstitials is further increased, the fit value of $\sigma^d_{xy}$ shifts to higher shear stress, and $\alpha$ decreases below one.
Fig.~\ref{fig:shear_trends}b shows the depinning stress as a function of interstitial number.
Fig.~\ref{fig:shear_trends}c shows the change in shear rate ($\kappa$) with shear stress as a function of interstitial number.
As more active interstitials are added, they not only increase the threshold of stress needed to de-pin dislocations, but also apply an additional drag to dislocation motion that persists after depinning.
The mechanism for this drag is subsequent re-acquisition of the moving stacking fault by active interstitial particles.

\subsection{Active vs. Passive Dislocation Pinning}

We have shown that stacking faults in crystals comprised of isotropically repulsive particles contain special voids that are favorable sites for rod-like interstitial particles to bind to.
When the stacking fault is associated with a partial dislocation, the migration of the dislocation core must be accompanied by the reconfiguration of this void, and therefore the ejection of the interstitial.
This ejection imposes a penalty on the shear force required to drive such dislocations into glide.
When these interstitial particles are active, they are capable of rapidly covering large distances in search of stacking faults.

This interaction of active interstitials and dislocations is interesting as an elaboration of the concept of a Cotrell atmosphere \cite{Cottrell1949}.
A Cottrell atmosphere is when interstitials collect around a dislocation core, effectively pinning its motion.
In atomic systems, interstitials are attracted to the dislocation because of the strain fields of both objects.
It is also possible for interstitials to have strong pinning interactions with the cores of dislocations \cite{Yu2015, Nabarro2004}.

Atomic dislocation-interstitial interactions have strong shear rate dependence \cite{Cottrell1949, Yoshinaga1971, Hirth1982, Fan2013} due to fundamental limitations of interstitial mobility.
Interstitials that interact strongly with the strained dislocation environment tend to also be slow diffusers, and so cannot quickly concentrate near or follow a moving dislocation.
Small interstitials that interact strongly with dislocation cores may be faster diffusers, however the core is a comparatively small region.
Therefore the probability of core-interstitial interaction depends heavily on the concentration of the solute species \cite{Nabarro2004}.

We find that active, rod-like interstitials avoid these problems by affecting greater decoupling between mobility and attraction to dislocations.
The active force allows such interstitials to cover large volumes compared to their passive counterparts.
Furthermore, active interstitials interact with stacking faults (not the dislocation core), which are extended defects.
This significantly improves the chance of interstitial-dislocation interception.
Our results show that active interstitials result in damped dislocation dynamics at shear rates much higher and concentrations lower than that at which passive interstitials would be effective.

\section{Conclusion} \label{section:conclusion}

We showed that rod-like active interstitials can interfere with dislocation motion at number concentrations as low as $64$ particles per million host particles.
This interference produces a shear stress threshold that was negligible without interstitials.
The physics behind the mechanism of binding and searching are antagonistic, so a compromise must be found for interstitial designs that strongly impact material plasticity.
We proposed a combined metric to determine active interstitials with desirable properties.
We found that the length of the rod-like interstitial is the dominant parameter for adjusting the rotational transition probability as well as binding preference in FCC crystals.
We expect that our findings on controlling plasticity in colloidal crystals using active interstitials will be useful in achieving deformable colloidal machines.

\section*{Acknowledgments}

This material is based upon work supported by the U.S. Department of Energy, Office of Science, Basic Energy Sciences, under Award Number DE-SC0013562 and by a University of Michigan Rackham Predoctoral Fellowship to B.V.S.
    This research utilized computational resources and services supported by Advanced Research Computing at the University of Michigan, Ann Arbor, and used the Extreme Science and Engineering Discovery Environment (XSEDE)\cite{XSEDE2014}, which is supported by National Science Foundation Grant ACI-1053575 (XSEDE Award DMR 140129).

\bibliography{rev}

\begin{thebibliography}{44}%
\makeatletter
\providecommand \@ifxundefined [1]{%
 \@ifx{#1\undefined}
}%
\providecommand \@ifnum [1]{%
 \ifnum #1\expandafter \@firstoftwo
 \else \expandafter \@secondoftwo
 \fi
}%
\providecommand \@ifx [1]{%
 \ifx #1\expandafter \@firstoftwo
 \else \expandafter \@secondoftwo
 \fi
}%
\providecommand \natexlab [1]{#1}%
\providecommand \enquote  [1]{``#1''}%
\providecommand \bibnamefont  [1]{#1}%
\providecommand \bibfnamefont [1]{#1}%
\providecommand \citenamefont [1]{#1}%
\providecommand \href@noop [0]{\@secondoftwo}%
\providecommand \href [0]{\begingroup \@sanitize@url \@href}%
\providecommand \@href[1]{\@@startlink{#1}\@@href}%
\providecommand \@@href[1]{\endgroup#1\@@endlink}%
\providecommand \@sanitize@url [0]{\catcode `\\12\catcode `\$12\catcode
  `\&12\catcode `\#12\catcode `\^12\catcode `\_12\catcode `\%12\relax}%
\providecommand \@@startlink[1]{}%
\providecommand \@@endlink[0]{}%
\providecommand \url  [0]{\begingroup\@sanitize@url \@url }%
\providecommand \@url [1]{\endgroup\@href {#1}{\urlprefix }}%
\providecommand \urlprefix  [0]{URL }%
\providecommand \Eprint [0]{\href }%
\providecommand \doibase [0]{http://dx.doi.org/}%
\providecommand \selectlanguage [0]{\@gobble}%
\providecommand \bibinfo  [0]{\@secondoftwo}%
\providecommand \bibfield  [0]{\@secondoftwo}%
\providecommand \translation [1]{[#1]}%
\providecommand \BibitemOpen [0]{}%
\providecommand \bibitemStop [0]{}%
\providecommand \bibitemNoStop [0]{.\EOS\space}%
\providecommand \EOS [0]{\spacefactor3000\relax}%
\providecommand \BibitemShut  [1]{\csname bibitem#1\endcsname}%
\let\auto@bib@innerbib\@empty
\bibitem [{\citenamefont {Li}\ \emph {et~al.}(2019)\citenamefont {Li},
  \citenamefont {Batra}, \citenamefont {Brown}, \citenamefont {Chang},
  \citenamefont {Ranganathan}, \citenamefont {Hoberman}, \citenamefont {Rus},\
  and\ \citenamefont {Lipson}}]{Li2019}%
  \BibitemOpen
  \bibfield  {author} {\bibinfo {author} {\bibfnamefont {S.}~\bibnamefont
  {Li}}, \bibinfo {author} {\bibfnamefont {R.}~\bibnamefont {Batra}}, \bibinfo
  {author} {\bibfnamefont {D.}~\bibnamefont {Brown}}, \bibinfo {author}
  {\bibfnamefont {H.-D.}\ \bibnamefont {Chang}}, \bibinfo {author}
  {\bibfnamefont {N.}~\bibnamefont {Ranganathan}}, \bibinfo {author}
  {\bibfnamefont {C.}~\bibnamefont {Hoberman}}, \bibinfo {author}
  {\bibfnamefont {D.}~\bibnamefont {Rus}}, \ and\ \bibinfo {author}
  {\bibfnamefont {H.}~\bibnamefont {Lipson}},\ }\href {\doibase
  10.1038/s41586-019-1022-9} {\bibfield  {journal} {\bibinfo  {journal}
  {Nature}\ }\textbf {\bibinfo {volume} {567}},\ \bibinfo {pages} {361}
  (\bibinfo {year} {2019})}\BibitemShut {NoStop}%
\bibitem [{\citenamefont {Koman}\ \emph {et~al.}(2018)\citenamefont {Koman},
  \citenamefont {Liu}, \citenamefont {Kozawa}, \citenamefont {Liu},
  \citenamefont {Cottrill}, \citenamefont {Son}, \citenamefont {Lebron},\ and\
  \citenamefont {Strano}}]{Koman2018}%
  \BibitemOpen
  \bibfield  {author} {\bibinfo {author} {\bibfnamefont {V.~B.}\ \bibnamefont
  {Koman}}, \bibinfo {author} {\bibfnamefont {P.}~\bibnamefont {Liu}}, \bibinfo
  {author} {\bibfnamefont {D.}~\bibnamefont {Kozawa}}, \bibinfo {author}
  {\bibfnamefont {A.~T.}\ \bibnamefont {Liu}}, \bibinfo {author} {\bibfnamefont
  {A.~L.}\ \bibnamefont {Cottrill}}, \bibinfo {author} {\bibfnamefont
  {Y.}~\bibnamefont {Son}}, \bibinfo {author} {\bibfnamefont {J.~A.}\
  \bibnamefont {Lebron}}, \ and\ \bibinfo {author} {\bibfnamefont {M.~S.}\
  \bibnamefont {Strano}},\ }\href {\doibase 10.1038/s41565-018-0194-z}
  {\bibfield  {journal} {\bibinfo  {journal} {Nature Nanotechnology}\ }\textbf
  {\bibinfo {volume} {13}},\ \bibinfo {pages} {819} (\bibinfo {year}
  {2018})}\BibitemShut {NoStop}%
\bibitem [{\citenamefont {Fletcher}\ and\ \citenamefont
  {Mullins}(2010)}]{Fletcher2010}%
  \BibitemOpen
  \bibfield  {author} {\bibinfo {author} {\bibfnamefont {D.~A.}\ \bibnamefont
  {Fletcher}}\ and\ \bibinfo {author} {\bibfnamefont {R.~D.}\ \bibnamefont
  {Mullins}},\ }\href {\doibase 10.1038/nature08908} {\bibfield  {journal}
  {\bibinfo  {journal} {Nature}\ }\textbf {\bibinfo {volume} {463}},\ \bibinfo
  {pages} {485} (\bibinfo {year} {2010})}\BibitemShut {NoStop}%
\bibitem [{\citenamefont {Rogers}\ \emph {et~al.}(2008)\citenamefont {Rogers},
  \citenamefont {Waigh},\ and\ \citenamefont {Lu}}]{Rogers2008}%
  \BibitemOpen
  \bibfield  {author} {\bibinfo {author} {\bibfnamefont {S.~S.}\ \bibnamefont
  {Rogers}}, \bibinfo {author} {\bibfnamefont {T.~A.}\ \bibnamefont {Waigh}}, \
  and\ \bibinfo {author} {\bibfnamefont {J.~R.}\ \bibnamefont {Lu}},\ }\href
  {\doibase 10.1529/biophysj.107.123851} {\bibfield  {journal} {\bibinfo
  {journal} {Biophysical Journal}\ }\textbf {\bibinfo {volume} {94}},\ \bibinfo
  {pages} {3313} (\bibinfo {year} {2008})}\BibitemShut {NoStop}%
\bibitem [{\citenamefont {Kim}\ \emph {et~al.}(2016)\citenamefont {Kim},
  \citenamefont {Macfarlane}, \citenamefont {Jones},\ and\ \citenamefont
  {Mirkin}}]{Kim2016}%
  \BibitemOpen
  \bibfield  {author} {\bibinfo {author} {\bibfnamefont {Y.}~\bibnamefont
  {Kim}}, \bibinfo {author} {\bibfnamefont {R.~J.}\ \bibnamefont {Macfarlane}},
  \bibinfo {author} {\bibfnamefont {M.~R.}\ \bibnamefont {Jones}}, \ and\
  \bibinfo {author} {\bibfnamefont {C.~A.}\ \bibnamefont {Mirkin}},\ }\href
  {\doibase 10.1126/science.aad2212} {\bibfield  {journal} {\bibinfo  {journal}
  {Science}\ }\textbf {\bibinfo {volume} {351}},\ \bibinfo {pages} {579}
  (\bibinfo {year} {2016})}\BibitemShut {NoStop}%
\bibitem [{\citenamefont {Yu}\ \emph {et~al.}(2018)\citenamefont {Yu},
  \citenamefont {Wang}, \citenamefont {Du}, \citenamefont {Wang},\ and\
  \citenamefont {Zhang}}]{Yu2018}%
  \BibitemOpen
  \bibfield  {author} {\bibinfo {author} {\bibfnamefont {J.}~\bibnamefont
  {Yu}}, \bibinfo {author} {\bibfnamefont {B.}~\bibnamefont {Wang}}, \bibinfo
  {author} {\bibfnamefont {X.}~\bibnamefont {Du}}, \bibinfo {author}
  {\bibfnamefont {Q.}~\bibnamefont {Wang}}, \ and\ \bibinfo {author}
  {\bibfnamefont {L.}~\bibnamefont {Zhang}},\ }\href {\doibase
  10.1038/s41467-018-05749-6} {\bibfield  {journal} {\bibinfo  {journal}
  {Nature Communications}\ }\textbf {\bibinfo {volume} {9}},\ \bibinfo {pages}
  {3260} (\bibinfo {year} {2018})}\BibitemShut {NoStop}%
\bibitem [{\citenamefont {Yigit}\ \emph {et~al.}(2019)\citenamefont {Yigit},
  \citenamefont {Alapan},\ and\ \citenamefont {Sitti}}]{Yigit2019}%
  \BibitemOpen
  \bibfield  {author} {\bibinfo {author} {\bibfnamefont {B.}~\bibnamefont
  {Yigit}}, \bibinfo {author} {\bibfnamefont {Y.}~\bibnamefont {Alapan}}, \
  and\ \bibinfo {author} {\bibfnamefont {M.}~\bibnamefont {Sitti}},\ }\href
  {\doibase 10.1002/advs.201801837} {\bibfield  {journal} {\bibinfo  {journal}
  {Advanced Science}\ }\textbf {\bibinfo {volume} {6}},\ \bibinfo {pages}
  {1801837} (\bibinfo {year} {2019})}\BibitemShut {NoStop}%
\bibitem [{\citenamefont {Xie}\ \emph {et~al.}(2019)\citenamefont {Xie},
  \citenamefont {Sun}, \citenamefont {Fan}, \citenamefont {Lin}, \citenamefont
  {Chen}, \citenamefont {Wang}, \citenamefont {Dong},\ and\ \citenamefont
  {He}}]{Xie2019}%
  \BibitemOpen
  \bibfield  {author} {\bibinfo {author} {\bibfnamefont {H.}~\bibnamefont
  {Xie}}, \bibinfo {author} {\bibfnamefont {M.}~\bibnamefont {Sun}}, \bibinfo
  {author} {\bibfnamefont {X.}~\bibnamefont {Fan}}, \bibinfo {author}
  {\bibfnamefont {Z.}~\bibnamefont {Lin}}, \bibinfo {author} {\bibfnamefont
  {W.}~\bibnamefont {Chen}}, \bibinfo {author} {\bibfnamefont {L.}~\bibnamefont
  {Wang}}, \bibinfo {author} {\bibfnamefont {L.}~\bibnamefont {Dong}}, \ and\
  \bibinfo {author} {\bibfnamefont {Q.}~\bibnamefont {He}},\ }\href {\doibase
  10.1126/scirobotics.aav8006} {\bibfield  {journal} {\bibinfo  {journal}
  {Science Robotics}\ }\textbf {\bibinfo {volume} {4}},\ \bibinfo {pages}
  {eaav8006} (\bibinfo {year} {2019})}\BibitemShut {NoStop}%
\bibitem [{\citenamefont {Hirth}\ and\ \citenamefont {{Lothe,
  Jens}}(1982)}]{Hirth1982}%
  \BibitemOpen
  \bibfield  {author} {\bibinfo {author} {\bibfnamefont {J.}~\bibnamefont
  {Hirth}}\ and\ \bibinfo {author} {\bibnamefont {{Lothe, Jens}}},\ }\href@noop
  {} {\emph {\bibinfo {title} {Theory of {Dislocations}}}},\ \bibinfo {edition}
  {2nd}\ ed.\ (\bibinfo  {publisher} {Wiley Interscience},\ \bibinfo {address}
  {New York},\ \bibinfo {year} {1982})\BibitemShut {NoStop}%
\bibitem [{\citenamefont {Schall}\ \emph {et~al.}(2004)\citenamefont {Schall},
  \citenamefont {Cohen}, \citenamefont {Weitz},\ and\ \citenamefont
  {Spaepen}}]{Schall2004}%
  \BibitemOpen
  \bibfield  {author} {\bibinfo {author} {\bibfnamefont {P.}~\bibnamefont
  {Schall}}, \bibinfo {author} {\bibfnamefont {I.}~\bibnamefont {Cohen}},
  \bibinfo {author} {\bibfnamefont {D.~A.}\ \bibnamefont {Weitz}}, \ and\
  \bibinfo {author} {\bibfnamefont {F.}~\bibnamefont {Spaepen}},\ }\href
  {\doibase 10.1126/science.1102186} {\bibfield  {journal} {\bibinfo  {journal}
  {Science}\ }\textbf {\bibinfo {volume} {305}},\ \bibinfo {pages} {1944}
  (\bibinfo {year} {2004})}\BibitemShut {NoStop}%
\bibitem [{\citenamefont {Schall}\ \emph {et~al.}(2006)\citenamefont {Schall},
  \citenamefont {Cohen}, \citenamefont {Weitz},\ and\ \citenamefont
  {Spaepen}}]{Schall2006}%
  \BibitemOpen
  \bibfield  {author} {\bibinfo {author} {\bibfnamefont {P.}~\bibnamefont
  {Schall}}, \bibinfo {author} {\bibfnamefont {I.}~\bibnamefont {Cohen}},
  \bibinfo {author} {\bibfnamefont {D.~A.}\ \bibnamefont {Weitz}}, \ and\
  \bibinfo {author} {\bibfnamefont {F.}~\bibnamefont {Spaepen}},\ }\href
  {\doibase 10.1038/nature04557} {\bibfield  {journal} {\bibinfo  {journal}
  {Nature}\ }\textbf {\bibinfo {volume} {440}},\ \bibinfo {pages} {319}
  (\bibinfo {year} {2006})}\BibitemShut {NoStop}%
\bibitem [{\citenamefont {Lin}\ \emph {et~al.}(2016)\citenamefont {Lin},
  \citenamefont {Bierbaum}, \citenamefont {Schall}, \citenamefont {Sethna},\
  and\ \citenamefont {Cohen}}]{Lin2016}%
  \BibitemOpen
  \bibfield  {author} {\bibinfo {author} {\bibfnamefont {N.~Y.~C.}\
  \bibnamefont {Lin}}, \bibinfo {author} {\bibfnamefont {M.}~\bibnamefont
  {Bierbaum}}, \bibinfo {author} {\bibfnamefont {P.}~\bibnamefont {Schall}},
  \bibinfo {author} {\bibfnamefont {J.~P.}\ \bibnamefont {Sethna}}, \ and\
  \bibinfo {author} {\bibfnamefont {I.}~\bibnamefont {Cohen}},\ }\href
  {\doibase 10.1038/nmat4715} {\bibfield  {journal} {\bibinfo  {journal}
  {Nature Materials}\ }\textbf {\bibinfo {volume} {15}},\ \bibinfo {pages}
  {1172} (\bibinfo {year} {2016})}\BibitemShut {NoStop}%
\bibitem [{\citenamefont {van~der Meer}\ \emph {et~al.}(2017)\citenamefont
  {van~der Meer}, \citenamefont {Dijkstra},\ and\ \citenamefont
  {Filion}}]{vdMeer2017}%
  \BibitemOpen
  \bibfield  {author} {\bibinfo {author} {\bibfnamefont {B.}~\bibnamefont
  {van~der Meer}}, \bibinfo {author} {\bibfnamefont {M.}~\bibnamefont
  {Dijkstra}}, \ and\ \bibinfo {author} {\bibfnamefont {L.}~\bibnamefont
  {Filion}},\ }\href {\doibase 10.1063/1.4990416} {\bibfield  {journal}
  {\bibinfo  {journal} {The Journal of Chemical Physics}\ }\textbf {\bibinfo
  {volume} {146}},\ \bibinfo {pages} {244905} (\bibinfo {year}
  {2017})}\BibitemShut {NoStop}%
\bibitem [{\citenamefont {VanSaders}\ \emph {et~al.}(2018)\citenamefont
  {VanSaders}, \citenamefont {Dshemuchadse},\ and\ \citenamefont
  {Glotzer}}]{VanSaders2018}%
  \BibitemOpen
  \bibfield  {author} {\bibinfo {author} {\bibfnamefont {B.}~\bibnamefont
  {VanSaders}}, \bibinfo {author} {\bibfnamefont {J.}~\bibnamefont
  {Dshemuchadse}}, \ and\ \bibinfo {author} {\bibfnamefont {S.~C.}\
  \bibnamefont {Glotzer}},\ }\href {\doibase 10.1103/PhysRevMaterials.2.063604}
  {\bibfield  {journal} {\bibinfo  {journal} {Physical Review Materials}\
  }\textbf {\bibinfo {volume} {2}},\ \bibinfo {pages} {063604} (\bibinfo {year}
  {2018})}\BibitemShut {NoStop}%
\bibitem [{\citenamefont {Glotzer}\ and\ \citenamefont
  {Solomon}(2007)}]{Glotzer2007}%
  \BibitemOpen
  \bibfield  {author} {\bibinfo {author} {\bibfnamefont {S.~C.}\ \bibnamefont
  {Glotzer}}\ and\ \bibinfo {author} {\bibfnamefont {M.~J.}\ \bibnamefont
  {Solomon}},\ }\href {\doibase 10.1038/nmat1949} {\bibfield  {journal}
  {\bibinfo  {journal} {Nature Materials}\ }\textbf {\bibinfo {volume} {6}},\
  \bibinfo {pages} {557} (\bibinfo {year} {2007})}\BibitemShut {NoStop}%
\bibitem [{\citenamefont {Sacanna}\ and\ \citenamefont
  {Pine}(2011)}]{Sacanna2011}%
  \BibitemOpen
  \bibfield  {author} {\bibinfo {author} {\bibfnamefont {S.}~\bibnamefont
  {Sacanna}}\ and\ \bibinfo {author} {\bibfnamefont {D.~J.}\ \bibnamefont
  {Pine}},\ }\href {\doibase 10.1016/j.cocis.2011.01.003} {\bibfield  {journal}
  {\bibinfo  {journal} {Current Opinion in Colloid \& Interface Science}\
  }\textbf {\bibinfo {volume} {16}},\ \bibinfo {pages} {96} (\bibinfo {year}
  {2011})}\BibitemShut {NoStop}%
\bibitem [{\citenamefont {Nykypanchuk}\ \emph {et~al.}(2008)\citenamefont
  {Nykypanchuk}, \citenamefont {Maye}, \citenamefont {Lelie},\ and\
  \citenamefont {Gang}}]{Nykypanchuk2008}%
  \BibitemOpen
  \bibfield  {author} {\bibinfo {author} {\bibfnamefont {D.}~\bibnamefont
  {Nykypanchuk}}, \bibinfo {author} {\bibfnamefont {M.~M.}\ \bibnamefont
  {Maye}}, \bibinfo {author} {\bibfnamefont {D.~v.~d.}\ \bibnamefont {Lelie}},
  \ and\ \bibinfo {author} {\bibfnamefont {O.}~\bibnamefont {Gang}},\ }\href
  {\doibase 10.1038/nature06560} {\bibfield  {journal} {\bibinfo  {journal}
  {Nature}\ }\textbf {\bibinfo {volume} {451}},\ \bibinfo {pages} {549}
  (\bibinfo {year} {2008})}\BibitemShut {NoStop}%
\bibitem [{\citenamefont {Boles}\ \emph {et~al.}(2016)\citenamefont {Boles},
  \citenamefont {Engel},\ and\ \citenamefont {Talapin}}]{Boles2016}%
  \BibitemOpen
  \bibfield  {author} {\bibinfo {author} {\bibfnamefont {M.~A.}\ \bibnamefont
  {Boles}}, \bibinfo {author} {\bibfnamefont {M.}~\bibnamefont {Engel}}, \ and\
  \bibinfo {author} {\bibfnamefont {D.~V.}\ \bibnamefont {Talapin}},\ }\href
  {\doibase 10.1021/acs.chemrev.6b00196} {\bibfield  {journal} {\bibinfo
  {journal} {Chemical Reviews}\ }\textbf {\bibinfo {volume} {116}},\ \bibinfo
  {pages} {11220} (\bibinfo {year} {2016})}\BibitemShut {NoStop}%
\bibitem [{\citenamefont {Marchetti}\ \emph {et~al.}(2013)\citenamefont
  {Marchetti}, \citenamefont {Joanny}, \citenamefont {Ramaswamy}, \citenamefont
  {Liverpool}, \citenamefont {Prost}, \citenamefont {Rao},\ and\ \citenamefont
  {Simha}}]{Marchetti2013}%
  \BibitemOpen
  \bibfield  {author} {\bibinfo {author} {\bibfnamefont {M.~C.}\ \bibnamefont
  {Marchetti}}, \bibinfo {author} {\bibfnamefont {J.~F.}\ \bibnamefont
  {Joanny}}, \bibinfo {author} {\bibfnamefont {S.}~\bibnamefont {Ramaswamy}},
  \bibinfo {author} {\bibfnamefont {T.~B.}\ \bibnamefont {Liverpool}}, \bibinfo
  {author} {\bibfnamefont {J.}~\bibnamefont {Prost}}, \bibinfo {author}
  {\bibfnamefont {M.}~\bibnamefont {Rao}}, \ and\ \bibinfo {author}
  {\bibfnamefont {R.~A.}\ \bibnamefont {Simha}},\ }\href {\doibase
  10.1103/RevModPhys.85.1143} {\bibfield  {journal} {\bibinfo  {journal}
  {Reviews of Modern Physics}\ }\textbf {\bibinfo {volume} {85}},\ \bibinfo
  {pages} {1143} (\bibinfo {year} {2013})}\BibitemShut {NoStop}%
\bibitem [{\citenamefont {Cottrell}\ and\ \citenamefont
  {Bilby}(1949)}]{Cottrell1949}%
  \BibitemOpen
  \bibfield  {author} {\bibinfo {author} {\bibfnamefont {A.~H.}\ \bibnamefont
  {Cottrell}}\ and\ \bibinfo {author} {\bibfnamefont {B.~A.}\ \bibnamefont
  {Bilby}},\ }\href {\doibase 10.1088/0370-1298/62/1/308} {\bibfield  {journal}
  {\bibinfo  {journal} {Proceedings of the Physical Society. Section A}\
  }\textbf {\bibinfo {volume} {62}},\ \bibinfo {pages} {49} (\bibinfo {year}
  {1949})}\BibitemShut {NoStop}%
\bibitem [{\citenamefont {Yoshinaga}\ and\ \citenamefont
  {Morozumi}(1971)}]{Yoshinaga1971}%
  \BibitemOpen
  \bibfield  {author} {\bibinfo {author} {\bibfnamefont {H.}~\bibnamefont
  {Yoshinaga}}\ and\ \bibinfo {author} {\bibfnamefont {S.}~\bibnamefont
  {Morozumi}},\ }\href {\doibase 10.1080/14786437108217008} {\bibfield
  {journal} {\bibinfo  {journal} {The Philosophical Magazine: A Journal of
  Theoretical Experimental and Applied Physics}\ }\textbf {\bibinfo {volume}
  {23}},\ \bibinfo {pages} {1367} (\bibinfo {year} {1971})}\BibitemShut
  {NoStop}%
\bibitem [{\citenamefont {Fan}\ \emph {et~al.}(2013)\citenamefont {Fan},
  \citenamefont {Osetskiy}, \citenamefont {Yip},\ and\ \citenamefont
  {Yildiz}}]{Fan2013}%
  \BibitemOpen
  \bibfield  {author} {\bibinfo {author} {\bibfnamefont {Y.}~\bibnamefont
  {Fan}}, \bibinfo {author} {\bibfnamefont {Y.~N.}\ \bibnamefont {Osetskiy}},
  \bibinfo {author} {\bibfnamefont {S.}~\bibnamefont {Yip}}, \ and\ \bibinfo
  {author} {\bibfnamefont {B.}~\bibnamefont {Yildiz}},\ }\href {\doibase
  10.1073/pnas.1310036110} {\bibfield  {journal} {\bibinfo  {journal}
  {Proceedings of the National Academy of Sciences}\ }\textbf {\bibinfo
  {volume} {110}},\ \bibinfo {pages} {17756} (\bibinfo {year}
  {2013})}\BibitemShut {NoStop}%
\bibitem [{\citenamefont {Weeks}\ \emph {et~al.}(1971)\citenamefont {Weeks},
  \citenamefont {Chandler},\ and\ \citenamefont {Andersen}}]{Weeks1971}%
  \BibitemOpen
  \bibfield  {author} {\bibinfo {author} {\bibfnamefont {J.}~\bibnamefont
  {Weeks}}, \bibinfo {author} {\bibfnamefont {D.}~\bibnamefont {Chandler}}, \
  and\ \bibinfo {author} {\bibfnamefont {H.~C.}\ \bibnamefont {Andersen}},\
  }\href {\doibase 10.1063/1.1674820} {\bibfield  {journal} {\bibinfo
  {journal} {The Journal of Chemical Physics}\ }\textbf {\bibinfo {volume}
  {54}},\ \bibinfo {pages} {5237} (\bibinfo {year} {1971})}\BibitemShut
  {NoStop}%
\bibitem [{\citenamefont {Pronk}\ and\ \citenamefont
  {Frenkel}(1999)}]{Pronk1999}%
  \BibitemOpen
  \bibfield  {author} {\bibinfo {author} {\bibfnamefont {S.}~\bibnamefont
  {Pronk}}\ and\ \bibinfo {author} {\bibfnamefont {D.}~\bibnamefont
  {Frenkel}},\ }\href {\doibase 10.1063/1.478339} {\bibfield  {journal}
  {\bibinfo  {journal} {The Journal of Chemical Physics}\ }\textbf {\bibinfo
  {volume} {110}},\ \bibinfo {pages} {4589} (\bibinfo {year}
  {1999})}\BibitemShut {NoStop}%
\bibitem [{\citenamefont {VanSaders}\ and\ \citenamefont
  {Glotzer}(2019)}]{VanSaders2019}%
  \BibitemOpen
  \bibfield  {author} {\bibinfo {author} {\bibfnamefont {B.}~\bibnamefont
  {VanSaders}}\ and\ \bibinfo {author} {\bibfnamefont {S.~C.}\ \bibnamefont
  {Glotzer}},\ }\href {\doibase 10.1039/C9SM00896A} {\bibfield  {journal}
  {\bibinfo  {journal} {Soft Matter}\ }\textbf {\bibinfo {volume} {15}},\
  \bibinfo {pages} {6086} (\bibinfo {year} {2019})}\BibitemShut {NoStop}%
\bibitem [{\citenamefont {Anderson}\ \emph {et~al.}(2008)\citenamefont
  {Anderson}, \citenamefont {Lorenz},\ and\ \citenamefont
  {Travesset}}]{Anderson2008}%
  \BibitemOpen
  \bibfield  {author} {\bibinfo {author} {\bibfnamefont {J.~A.}\ \bibnamefont
  {Anderson}}, \bibinfo {author} {\bibfnamefont {C.~D.}\ \bibnamefont
  {Lorenz}}, \ and\ \bibinfo {author} {\bibfnamefont {A.}~\bibnamefont
  {Travesset}},\ }\href {\doibase 10.1016/j.jcp.2008.01.047} {\bibfield
  {journal} {\bibinfo  {journal} {Journal of Computational Physics}\ }\textbf
  {\bibinfo {volume} {227}},\ \bibinfo {pages} {5342} (\bibinfo {year}
  {2008})}\BibitemShut {NoStop}%
\bibitem [{\citenamefont {Glaser}\ \emph {et~al.}(2015)\citenamefont {Glaser},
  \citenamefont {Nguyen}, \citenamefont {Anderson}, \citenamefont {Lui},
  \citenamefont {Spiga}, \citenamefont {Millan}, \citenamefont {Morse},\ and\
  \citenamefont {Glotzer}}]{Glaser2015}%
  \BibitemOpen
  \bibfield  {author} {\bibinfo {author} {\bibfnamefont {J.}~\bibnamefont
  {Glaser}}, \bibinfo {author} {\bibfnamefont {T.~D.}\ \bibnamefont {Nguyen}},
  \bibinfo {author} {\bibfnamefont {J.~A.}\ \bibnamefont {Anderson}}, \bibinfo
  {author} {\bibfnamefont {P.}~\bibnamefont {Lui}}, \bibinfo {author}
  {\bibfnamefont {F.}~\bibnamefont {Spiga}}, \bibinfo {author} {\bibfnamefont
  {J.~A.}\ \bibnamefont {Millan}}, \bibinfo {author} {\bibfnamefont {D.~C.}\
  \bibnamefont {Morse}}, \ and\ \bibinfo {author} {\bibfnamefont {S.~C.}\
  \bibnamefont {Glotzer}},\ }\href {\doibase 10.1016/j.cpc.2015.02.028}
  {\bibfield  {journal} {\bibinfo  {journal} {Computer Physics Communications}\
  }\textbf {\bibinfo {volume} {192}},\ \bibinfo {pages} {97} (\bibinfo {year}
  {2015})}\BibitemShut {NoStop}%
\bibitem [{\citenamefont {Martyna}\ \emph {et~al.}(1996)\citenamefont
  {Martyna}, \citenamefont {Tuckerman}, \citenamefont {Tobias},\ and\
  \citenamefont {Klein}}]{Martyna1996}%
  \BibitemOpen
  \bibfield  {author} {\bibinfo {author} {\bibfnamefont {G.~J.}\ \bibnamefont
  {Martyna}}, \bibinfo {author} {\bibfnamefont {M.~E.}\ \bibnamefont
  {Tuckerman}}, \bibinfo {author} {\bibfnamefont {D.~J.}\ \bibnamefont
  {Tobias}}, \ and\ \bibinfo {author} {\bibfnamefont {M.~L.}\ \bibnamefont
  {Klein}},\ }\href {\doibase 10.1080/00268979600100761} {\bibfield  {journal}
  {\bibinfo  {journal} {Molecular Physics}\ }\textbf {\bibinfo {volume} {87}},\
  \bibinfo {pages} {1117} (\bibinfo {year} {1996})}\BibitemShut {NoStop}%
\bibitem [{\citenamefont {Filion}\ \emph {et~al.}(2011)\citenamefont {Filion},
  \citenamefont {Ni}, \citenamefont {Frenkel},\ and\ \citenamefont
  {Dijkstra}}]{Filion2011}%
  \BibitemOpen
  \bibfield  {author} {\bibinfo {author} {\bibfnamefont {L.}~\bibnamefont
  {Filion}}, \bibinfo {author} {\bibfnamefont {R.}~\bibnamefont {Ni}}, \bibinfo
  {author} {\bibfnamefont {D.}~\bibnamefont {Frenkel}}, \ and\ \bibinfo
  {author} {\bibfnamefont {M.}~\bibnamefont {Dijkstra}},\ }\href {\doibase
  10.1063/1.3572059} {\bibfield  {journal} {\bibinfo  {journal} {The Journal of
  Chemical Physics}\ }\textbf {\bibinfo {volume} {134}},\ \bibinfo {pages}
  {134901} (\bibinfo {year} {2011})}\BibitemShut {NoStop}%
\bibitem [{\citenamefont {Nguyen}\ \emph {et~al.}(2011)\citenamefont {Nguyen},
  \citenamefont {Phillips}, \citenamefont {Anderson},\ and\ \citenamefont
  {Glotzer}}]{Nguyen2011}%
  \BibitemOpen
  \bibfield  {author} {\bibinfo {author} {\bibfnamefont {T.~D.}\ \bibnamefont
  {Nguyen}}, \bibinfo {author} {\bibfnamefont {C.~L.}\ \bibnamefont
  {Phillips}}, \bibinfo {author} {\bibfnamefont {J.~A.}\ \bibnamefont
  {Anderson}}, \ and\ \bibinfo {author} {\bibfnamefont {S.~C.}\ \bibnamefont
  {Glotzer}},\ }\href {\doibase 10.1016/j.cpc.2011.06.005} {\bibfield
  {journal} {\bibinfo  {journal} {Computer Physics Communications}\ }\textbf
  {\bibinfo {volume} {182}},\ \bibinfo {pages} {2307} (\bibinfo {year}
  {2011})}\BibitemShut {NoStop}%
\bibitem [{\citenamefont {Cazals}\ and\ \citenamefont
  {Dreyfus}(2017)}]{Cazals2017}%
  \BibitemOpen
  \bibfield  {author} {\bibinfo {author} {\bibfnamefont {F.}~\bibnamefont
  {Cazals}}\ and\ \bibinfo {author} {\bibfnamefont {T.}~\bibnamefont
  {Dreyfus}},\ }\href {\doibase 10.1093/bioinformatics/btw752} {\bibfield
  {journal} {\bibinfo  {journal} {Bioinformatics}\ }\textbf {\bibinfo {volume}
  {33}},\ \bibinfo {pages} {997} (\bibinfo {year} {2017})}\BibitemShut
  {NoStop}%
\bibitem [{\citenamefont {Parrinello}\ and\ \citenamefont
  {Rahman}(1982)}]{Parrinello1982}%
  \BibitemOpen
  \bibfield  {author} {\bibinfo {author} {\bibfnamefont {M.}~\bibnamefont
  {Parrinello}}\ and\ \bibinfo {author} {\bibfnamefont {A.}~\bibnamefont
  {Rahman}},\ }\href {\doibase http://dx.doi.org/10.1063/1.443248} {\bibfield
  {journal} {\bibinfo  {journal} {The Journal of Chemical Physics}\ }\textbf
  {\bibinfo {volume} {76}},\ \bibinfo {pages} {2662} (\bibinfo {year}
  {1982})}\BibitemShut {NoStop}%
\bibitem [{\citenamefont {Chandler}(1985)}]{Chandler1985}%
  \BibitemOpen
  \bibfield  {author} {\bibinfo {author} {\bibfnamefont {D.}~\bibnamefont
  {Chandler}},\ }\href@noop {} {\emph {\bibinfo {title} {Introduction to modern
  statistical mechanics}}}\ (\bibinfo  {publisher} {Oxford University Press},\
  \bibinfo {address} {New York},\ \bibinfo {year} {1985})\BibitemShut {NoStop}%
\bibitem [{\citenamefont {Viswanathan}\ \emph {et~al.}(1999)\citenamefont
  {Viswanathan}, \citenamefont {Buldyrev}, \citenamefont {Havlin},
  \citenamefont {da~Luz}, \citenamefont {Raposo},\ and\ \citenamefont
  {Stanley}}]{Viswanathan1999}%
  \BibitemOpen
  \bibfield  {author} {\bibinfo {author} {\bibfnamefont {G.~M.}\ \bibnamefont
  {Viswanathan}}, \bibinfo {author} {\bibfnamefont {S.~V.}\ \bibnamefont
  {Buldyrev}}, \bibinfo {author} {\bibfnamefont {S.}~\bibnamefont {Havlin}},
  \bibinfo {author} {\bibfnamefont {M.~G.~E.}\ \bibnamefont {da~Luz}}, \bibinfo
  {author} {\bibfnamefont {E.~P.}\ \bibnamefont {Raposo}}, \ and\ \bibinfo
  {author} {\bibfnamefont {H.~E.}\ \bibnamefont {Stanley}},\ }\href {\doibase
  10.1038/44831} {\bibfield  {journal} {\bibinfo  {journal} {Nature}\ }\textbf
  {\bibinfo {volume} {401}},\ \bibinfo {pages} {911} (\bibinfo {year}
  {1999})}\BibitemShut {NoStop}%
\bibitem [{\citenamefont {Bartumeus}\ \emph {et~al.}(2003)\citenamefont
  {Bartumeus}, \citenamefont {Peters}, \citenamefont {Pueyo}, \citenamefont
  {Marrasé},\ and\ \citenamefont {Catalan}}]{Bartumeus2003}%
  \BibitemOpen
  \bibfield  {author} {\bibinfo {author} {\bibfnamefont {F.}~\bibnamefont
  {Bartumeus}}, \bibinfo {author} {\bibfnamefont {F.}~\bibnamefont {Peters}},
  \bibinfo {author} {\bibfnamefont {S.}~\bibnamefont {Pueyo}}, \bibinfo
  {author} {\bibfnamefont {C.}~\bibnamefont {Marrasé}}, \ and\ \bibinfo
  {author} {\bibfnamefont {J.}~\bibnamefont {Catalan}},\ }\href {\doibase
  10.1073/pnas.2137243100} {\bibfield  {journal} {\bibinfo  {journal}
  {Proceedings of the National Academy of Sciences}\ }\textbf {\bibinfo
  {volume} {100}},\ \bibinfo {pages} {12771} (\bibinfo {year}
  {2003})}\BibitemShut {NoStop}%
\bibitem [{\citenamefont {Reynolds}\ and\ \citenamefont
  {Rhodes}(2009)}]{Reynolds2009}%
  \BibitemOpen
  \bibfield  {author} {\bibinfo {author} {\bibfnamefont {A.~M.}\ \bibnamefont
  {Reynolds}}\ and\ \bibinfo {author} {\bibfnamefont {C.~J.}\ \bibnamefont
  {Rhodes}},\ }\href {\doibase 10.1890/08-0153.1} {\bibfield  {journal}
  {\bibinfo  {journal} {Ecology}\ }\textbf {\bibinfo {volume} {90}},\ \bibinfo
  {pages} {877} (\bibinfo {year} {2009})}\BibitemShut {NoStop}%
\bibitem [{\citenamefont {Humphries}\ \emph {et~al.}(2012)\citenamefont
  {Humphries}, \citenamefont {Weimerskirch}, \citenamefont {Queiroz},
  \citenamefont {Southall},\ and\ \citenamefont {Sims}}]{Humphries2012}%
  \BibitemOpen
  \bibfield  {author} {\bibinfo {author} {\bibfnamefont {N.~E.}\ \bibnamefont
  {Humphries}}, \bibinfo {author} {\bibfnamefont {H.}~\bibnamefont
  {Weimerskirch}}, \bibinfo {author} {\bibfnamefont {N.}~\bibnamefont
  {Queiroz}}, \bibinfo {author} {\bibfnamefont {E.~J.}\ \bibnamefont
  {Southall}}, \ and\ \bibinfo {author} {\bibfnamefont {D.~W.}\ \bibnamefont
  {Sims}},\ }\href {\doibase 10.1073/pnas.1121201109} {\bibfield  {journal}
  {\bibinfo  {journal} {Proceedings of the National Academy of Sciences}\
  }\textbf {\bibinfo {volume} {109}},\ \bibinfo {pages} {7169} (\bibinfo {year}
  {2012})}\BibitemShut {NoStop}%
\bibitem [{\citenamefont {Haklı}\ and\ \citenamefont
  {Uğuz}(2014)}]{Hakli2014}%
  \BibitemOpen
  \bibfield  {author} {\bibinfo {author} {\bibfnamefont {H.}~\bibnamefont
  {Haklı}}\ and\ \bibinfo {author} {\bibfnamefont {H.}~\bibnamefont {Uğuz}},\
  }\href {\doibase 10.1016/j.asoc.2014.06.034} {\bibfield  {journal} {\bibinfo
  {journal} {Applied Soft Computing}\ }\textbf {\bibinfo {volume} {23}},\
  \bibinfo {pages} {333} (\bibinfo {year} {2014})}\BibitemShut {NoStop}%
\bibitem [{\citenamefont {Rupprecht}\ \emph {et~al.}(2016)\citenamefont
  {Rupprecht}, \citenamefont {Bénichou},\ and\ \citenamefont
  {Voituriez}}]{Rupprecht2016}%
  \BibitemOpen
  \bibfield  {author} {\bibinfo {author} {\bibfnamefont {J.-F.}\ \bibnamefont
  {Rupprecht}}, \bibinfo {author} {\bibfnamefont {O.}~\bibnamefont
  {Bénichou}}, \ and\ \bibinfo {author} {\bibfnamefont {R.}~\bibnamefont
  {Voituriez}},\ }\href {\doibase 10.1103/PhysRevE.94.012117} {\bibfield
  {journal} {\bibinfo  {journal} {Physical Review E}\ }\textbf {\bibinfo
  {volume} {94}},\ \bibinfo {pages} {012117} (\bibinfo {year}
  {2016})}\BibitemShut {NoStop}%
\bibitem [{\citenamefont {Stukowski}(2010)}]{OVITO}%
  \BibitemOpen
  \bibfield  {author} {\bibinfo {author} {\bibfnamefont {A.}~\bibnamefont
  {Stukowski}},\ }\href {\doibase 10.1088/0965-0393/18/1/015012} {\bibfield
  {journal} {\bibinfo  {journal} {Modelling and Simulation in Materials Science
  and Engineering}\ }\textbf {\bibinfo {volume} {18}},\ \bibinfo {pages}
  {015012} (\bibinfo {year} {2010})}\BibitemShut {NoStop}%
\bibitem [{\citenamefont {Bhattacharjee}\ and\ \citenamefont
  {Datta}(2019)}]{Bhattacharjee2019}%
  \BibitemOpen
  \bibfield  {author} {\bibinfo {author} {\bibfnamefont {T.}~\bibnamefont
  {Bhattacharjee}}\ and\ \bibinfo {author} {\bibfnamefont {S.~S.}\ \bibnamefont
  {Datta}},\ }\href {\doibase 10.1039/C9SM01735F} {\bibfield  {journal}
  {\bibinfo  {journal} {Soft Matter}\ }\textbf {\bibinfo {volume} {15}},\
  \bibinfo {pages} {9920} (\bibinfo {year} {2019})}\BibitemShut {NoStop}%
\bibitem [{\citenamefont {Yu}\ \emph {et~al.}(2015)\citenamefont {Yu},
  \citenamefont {Qi}, \citenamefont {Tsuru}, \citenamefont {Traylor},
  \citenamefont {Rugg}, \citenamefont {Morris}, \citenamefont {Asta},
  \citenamefont {Chrzan},\ and\ \citenamefont {Minor}}]{Yu2015}%
  \BibitemOpen
  \bibfield  {author} {\bibinfo {author} {\bibfnamefont {Q.}~\bibnamefont
  {Yu}}, \bibinfo {author} {\bibfnamefont {L.}~\bibnamefont {Qi}}, \bibinfo
  {author} {\bibfnamefont {T.}~\bibnamefont {Tsuru}}, \bibinfo {author}
  {\bibfnamefont {R.}~\bibnamefont {Traylor}}, \bibinfo {author} {\bibfnamefont
  {D.}~\bibnamefont {Rugg}}, \bibinfo {author} {\bibfnamefont {J.~W.}\
  \bibnamefont {Morris}}, \bibinfo {author} {\bibfnamefont {M.}~\bibnamefont
  {Asta}}, \bibinfo {author} {\bibfnamefont {D.~C.}\ \bibnamefont {Chrzan}}, \
  and\ \bibinfo {author} {\bibfnamefont {A.~M.}\ \bibnamefont {Minor}},\ }\href
  {\doibase 10.1126/science.1260485} {\bibfield  {journal} {\bibinfo  {journal}
  {Science}\ }\textbf {\bibinfo {volume} {347}},\ \bibinfo {pages} {635}
  (\bibinfo {year} {2015})}\BibitemShut {NoStop}%
\bibitem [{\citenamefont {Nabarro}\ and\ \citenamefont
  {Hirth}(2004)}]{Nabarro2004}%
  \BibitemOpen
  \bibinfo {editor} {\bibfnamefont {F.~R.~N.}\ \bibnamefont {Nabarro}}\ and\
  \bibinfo {editor} {\bibfnamefont {J.~P.}\ \bibnamefont {Hirth}},\ eds.,\
  \href@noop {} {\emph {\bibinfo {title} {Dislocations in {Solids}}}},\
  \bibinfo {edition} {1st}\ ed.,\ Vol.~\bibinfo {volume} {12}\ (\bibinfo
  {publisher} {North Holland},\ \bibinfo {year} {2004})\BibitemShut {NoStop}%
\bibitem [{\citenamefont {Towns}\ \emph {et~al.}(2014)\citenamefont {Towns},
  \citenamefont {Cockerill}, \citenamefont {Dahan}, \citenamefont {Foster},
  \citenamefont {Gaither}, \citenamefont {Grimshaw}, \citenamefont {Hazlewood},
  \citenamefont {Lathrop}, \citenamefont {Lifka}, \citenamefont {Peterson},
  \citenamefont {Roskies}, \citenamefont {Scott},\ and\ \citenamefont
  {Wilkins-Diehr}}]{XSEDE2014}%
  \BibitemOpen
  \bibfield  {author} {\bibinfo {author} {\bibfnamefont {J.}~\bibnamefont
  {Towns}}, \bibinfo {author} {\bibfnamefont {T.}~\bibnamefont {Cockerill}},
  \bibinfo {author} {\bibfnamefont {M.}~\bibnamefont {Dahan}}, \bibinfo
  {author} {\bibfnamefont {I.}~\bibnamefont {Foster}}, \bibinfo {author}
  {\bibfnamefont {K.}~\bibnamefont {Gaither}}, \bibinfo {author} {\bibfnamefont
  {A.}~\bibnamefont {Grimshaw}}, \bibinfo {author} {\bibfnamefont
  {V.}~\bibnamefont {Hazlewood}}, \bibinfo {author} {\bibfnamefont
  {S.}~\bibnamefont {Lathrop}}, \bibinfo {author} {\bibfnamefont
  {D.}~\bibnamefont {Lifka}}, \bibinfo {author} {\bibfnamefont {G.~D.}\
  \bibnamefont {Peterson}}, \bibinfo {author} {\bibfnamefont {R.}~\bibnamefont
  {Roskies}}, \bibinfo {author} {\bibfnamefont {J.~R.}\ \bibnamefont {Scott}},
  \ and\ \bibinfo {author} {\bibfnamefont {N.}~\bibnamefont {Wilkins-Diehr}},\
  }\href {\doibase 10.1109/MCSE.2014.80} {\bibfield  {journal} {\bibinfo
  {journal} {Computing in Science Engineering}\ }\textbf {\bibinfo {volume}
  {16}},\ \bibinfo {pages} {62} (\bibinfo {year} {2014})}\BibitemShut {NoStop}%
\end{thebibliography}%


\providecommand*{\mcitethebibliography}{\thebibliography}
\csname @ifundefined\endcsname{endmcitethebibliography}
{\let\endmcitethebibliography\endthebibliography}{}

\end{document}